\documentclass[11pt, a4paper]{article}
\usepackage{jcappub}
\usepackage{siunitx}
\DeclareSIUnit\parsec{pc}
\DeclareSIUnit\dex{dex}
\usepackage{collcell}
\usepackage{afterpage}
\usepackage{layouts}
\usepackage[version=4]{mhchem}
\usepackage{tikz}
\usetikzlibrary{patterns}
\usetikzlibrary{decorations.pathmorphing}
\usetikzlibrary{decorations.markings}
\usetikzlibrary{angles,quotes}
\tikzset{photon/.style={decorate, decoration={snake}, draw=red}}
\usepackage{cleveref}
\crefname{equation}{Eq.}{Eqs.}
\Crefname{equation}{Equation}{Equations}
\crefname{figure}{Fig.}{Figs.}
\Crefname{figure}{Figure}{Figures}
\crefname{table}{Table}{Tables}
\crefname{section}{Section}{Sections}
\crefname{appendix}{Appendix}{Appendices}

\pdfoutput=1

\usepackage[T1]{fontenc}
\usepackage{lmodern}
\usepackage{calc}
\usepackage{graphicx}
\usepackage{booktabs}
\usepackage{textcomp}
\usepackage{xspace}
\usepackage{relsize}
\usepackage{amssymb}
\usepackage{amsmath}
\usepackage{listings}
\usepackage{microtype}
\usepackage{multirow}
\usepackage{tabularx}
\usepackage{array}
\usepackage{placeins}
\usepackage{cuted}
\usepackage{soul} 
\usepackage{fixltx2e}
\usepackage{slashed}
\usepackage{bm}
\usepackage[numbers,sort&compress]{natbib}
\usepackage[labelfont=bf,font=small]{caption}
\usepackage[skip=-2pt]{subcaption}
\usepackage[clockwise,figuresright]{rotating}
\usepackage{tikz}
\usepackage[normalem]{ulem}
\usepackage[utf8]{inputenc}

\usepackage{etoolbox}
\AfterEndEnvironment{strip}{\leavevmode}

\allowdisplaybreaks

\newcolumntype{L}{>{\raggedright\let\newline\\\arraybackslash\hspace{0pt}}X}
\newcolumntype{R}{>{\raggedleft\let\newline\\\arraybackslash\hspace{0pt}}X}
\newcolumntype{C}{>{\centering\let\newline\\\arraybackslash\hspace{0pt}}X}

\newcommand{\gambitinstitute}[1]{\expandafter\csname #1\endcsname\label{#1}}

\newcommand{\aachen}{Institute for Theoretical Particle Physics and Cosmology (TTK), RWTH Aachen University, D-52056 Aachen, Germany}

\newcommand{\imperial}{Department of Physics, Imperial College London, Blackett Laboratory, Prince Consort Road, London SW7 2AZ, UK}
\newcommand{\cambridge}{Cavendish Laboratory, University of Cambridge, JJ Thomson Avenue, Cambridge, CB3 0HE, UK}

\newcommand{\monash}{School of Physics and Astronomy, Monash University, Melbourne, VIC 3800, Australia}

\newcommand{\okc}{Oskar Klein Centre for Cosmoparticle Physics, AlbaNova University Centre, SE-10691 Stockholm, Sweden}

\newcommand{\iap}{Institut d'Astrophysique de Paris, 98~bis boulevard Arago, 75014 Paris, France}

\newcommand{\uq}{School of Mathematics and Physics, The University of Queensland, St.\ Lucia, Brisbane, QLD 4072, Australia}
\newcommand{\gottingen}{Institut f\"ur Astrophysik und Geophysik, Georg-August-Universit\"at G\"ottingen, Friedrich-Hund-Platz~1, 37077 G\"ottingen, Germany}

\newcommand{\kicc}{Kavli Institute for Cosmology, University of Cambridge, Madingley Road, Cambridge, CB3 0HA, UK}

\newcommand{\kings}{Theoretical Particle Physics and Cosmology Group, Department of Physics, King’s College London, Strand, London, WC2R 2LS, UK}

\newcommand{\kitIAP}{Institute for Astroparticle Physics (IAP), Karlsruhe Institute of Technology (KIT), Hermann-von-Helmholtz-Platz 1, D-76344 Eggenstein-Leopoldshafen, Germany}
\newcommand{\kitTTP}{Institute for Theoretical Particle Physics (TTP), Karlsruhe Institute of Technology (KIT), 76128 Karlsruhe, Germany}

\makeatletter

\newcommand{\preprintnumber}[1]{\gdef\@preprintnumber{\begin{flushright}{#1}\end{flushright}}}

\g@addto@macro\bfseries{\boldmath}
\makeatother

\let\underscore\_
\renewcommand{\_}{\discretionary{\underscore}{}{\underscore}}

\makeatletter
\let\orgdescriptionlabel\descriptionlabel
\renewcommand*{\descriptionlabel}[1]{%
  \let\orglabel\label
  \let\label\@gobble
  \phantomsection
  \protected@edef\@currentlabel{#1}%
  \let\label\orglabel
  \orgdescriptionlabel{#1}%
}
\makeatother

\newcommand\cpp[1]{{\lstinline!#1!}}  

\newcommand\yaml[1]{{\lstset{style=yaml}\lstinline!#1!\lstset{style=cpp}}}

\newcommand\term[1]{{\lstset{style=terminal}\lstinline!#1!\lstset{style=cpp}}}
\newcommand\termalt[1]{{\lstset{style=terminalalt}\lstinline!#1!\lstset{style=cpp}}}
\newcommand\fortran[1]{{\lstset{style=fortran}\lstinline!#1!\lstset{style=cpp}}}
\newcommand\py[1]{{\lstset{style=python}\lstinline!#1!\lstset{style=cpp}}}
\newcommand\customtilde{{\raisebox{0.2ex}{\scalebox{0.6}{\boldmath$\sim$}}}}
\newcommand\mathematica[1]{{\lstset{style=Mathematica}\lstinline!#1!\lstset{style=cpp}}}
\newcommand\guminline[1]{{{\lstset{style=gum}\lstinline!#1!}}}
\newcommand\textinline[1]{{{\lstset{style=text}\lstinline!#1!}}}

\def\be{\begin{equation}}
\def\ee{\end{equation}}
\def\ba{\begin{eqnarray}}
\def\ea{\end{eqnarray}}
\newcommand{\bea}{\begin{eqnarray}}
\newcommand{\eea}{\end{eqnarray}}

\lstnewenvironment{lstlistingyaml}{\lstset{style=yaml}}{\lstset{style=cpp}}
\lstnewenvironment{lstlistingterm}{\lstset{style=terminal}}{\lstset{style=cpp}}
\lstnewenvironment{lstlistingfortran}{\lstset{style=fortran}}{\lstset{style=cpp}}
\lstnewenvironment{lstcpp}{\lstset{style=cpp}}{\lstset{style=cpp}}
\lstnewenvironment{lstcppalt}{\lstset{style=cppalt}}{\lstset{style=cpp}}
\lstnewenvironment{lstcppnum}{\lstset{style=cppnum}}{\lstset{style=cpp}}
\lstnewenvironment{lstyaml}{\lstset{style=yaml}}{\lstset{style=cpp}}
\lstnewenvironment{lstgum}{\lstset{style=gum}}{\lstset{style=cpp}}
\lstnewenvironment{lstterm}{\lstset{style=terminal}}{\lstset{style=cpp}}
\lstnewenvironment{lsttermalt}{\lstset{style=terminalalt}}{\lstset{style=cpp}}
\lstnewenvironment{lsttext}{\lstset{style=text}}{\lstset{style=cpp}}
\lstnewenvironment{lstfortran}{\lstset{style=fortran}}{\lstset{style=cpp}}
\lstnewenvironment{lstpy}{\lstset{style=python}}{\lstset{style=cpp}}
\lstnewenvironment{lstmathematica}{\lstset{style=mathematica}}{\lstset{style=cpp}}

\newcommand{\tmpname}{}
\newcommand{\tmplistingname}{}
\makeatletter
\newif\ifATOlabelname
\lst@Key{labelname}{Listing}{\def\ATOlabelname{#1}\global\ATOlabelnametrue}
\makeatother
\lstnewenvironment{lstcpplabel}[1][]{
  \lstset{style=cpp,#1} 
  \ifATOlabelname
    \renewcommand{\tmpname}{\lstlistingname}
    \renewcommand{\tmplistingname}{\lstlistlistingname}
    \renewcommand{\lstlistingname}{\ATOlabelname}
    \renewcommand{\lstlistlistingname}{List of \lstlistingname s}
  \fi
}{
  \renewcommand{\lstlistingname}{\tmpname}
  \renewcommand{\lstlistlistingname}{\tmplistingname}
  \lstset{style=cpp}
}
\definecolor{solarized@base03}{HTML}{002B36}
\definecolor{solarized@base02}{HTML}{073642}
\definecolor{solarized@base01}{HTML}{586e75}
\definecolor{solarized@base00}{HTML}{657b83}
\definecolor{solarized@base0}{HTML}{839496}
\definecolor{solarized@base1}{HTML}{93a1a1}
\definecolor{solarized@base2}{HTML}{EEE8D5}
\definecolor{solarized@base3}{HTML}{FDF6E3}
\definecolor{solarized@yellow}{HTML}{B58900}
\definecolor{solarized@orange}{HTML}{CB4B16}
\definecolor{solarized@red}{HTML}{DC322F}
\definecolor{solarized@magenta}{HTML}{D33682}
\definecolor{solarized@violet}{HTML}{6C71C4}
\definecolor{solarized@blue}{HTML}{268BD2}
\definecolor{solarized@cyan}{HTML}{2AA198}
\definecolor{solarized@green}{HTML}{859900}
\definecolor{darkred}{HTML}{550003}
\definecolor{darkgreen}{HTML}{00AA00}
\definecolor{orchid}{HTML}{AF06F5}

\newcommand\YAMLstringstyle{\footnotesize\color{solarized@green}\mdseries}
\newcommand\YAMLkeystyle{\footnotesize\color{solarized@blue}\ttfamily}
\newcommand\YAMLvaluestyle{\footnotesize\color{blue}\mdseries}
\newcommand\ProcessThreeDashes{\llap{\color{cyan}\mdseries-{-}-}}

\newcommand\CPPcommentstyle{\color{solarized@violet}\footnotesize\ttfamily}
\newcommand\CPPdirectivestyle{\color{solarized@magenta}\footnotesize\ttfamily}
\newcommand\termplainstyle{\footnotesize\ttfamily}

\newcommand\YAMLcommentstyle{\color{solarized@orange}\ttfamily}

\newcommand\processLongMacroDelimiter
{%
\CPPdirectivestyle%
\#define%
}

\lstdefinestyle{cpp}
{
  language=C++,
  basicstyle=\footnotesize\ttfamily,
  basewidth={0.53em,0.44em}, 
  numbers=none,
  tabsize=2,
  breaklines=true,
  escapeinside={@}{@},
  showstringspaces=false,
  numberstyle=\tiny\color{solarized@base01},
  keywordstyle=\color{solarized@orange},
  stringstyle=\color{solarized@red}\ttfamily,
  identifierstyle=\color{solarized@blue},
  commentstyle=\CPPcommentstyle,
  directivestyle=\CPPdirectivestyle,
  emphstyle=\color{solarized@green},
  frame=single,
  rulecolor=\color{solarized@base2},
  rulesepcolor=\color{solarized@base2},
  literate={~} {\customtilde}1,
  moredelim=*[directive]\ \ \#,
  moredelim=*[directive]\ \ \ \ \#
}

\lstdefinestyle{cppalt}
{
  language=C++,
  basicstyle=\footnotesize\ttfamily,
  basewidth={0.53em,0.44em}, 
  numbers=none,
  tabsize=2,
  breaklines=true,
  escapeinside={*@}{@*},
  showstringspaces=false,
  numberstyle=\tiny\color{solarized@base01},
  keywordstyle=\color{solarized@orange},
  stringstyle=\color{solarized@red}\ttfamily,
  identifierstyle=\color{solarized@blue},
  commentstyle=\CPPcommentstyle,
  directivestyle=\CPPdirectivestyle,
  emphstyle=\color{solarized@green},
  frame=single,
  rulecolor=\color{solarized@base2},
  rulesepcolor=\color{solarized@base2},
  literate={~}{\customtilde}1,
  moredelim=**[is][\processLongMacroDelimiter]{BeginLongMacro}{EndLongMacro} 
}

\lstdefinestyle{cppnum}
{
  language=C++,
  basicstyle=\footnotesize\ttfamily,
  basewidth={0.53em,0.44em}, 
  numbers=none,
  tabsize=2,
  breaklines=true,
  escapeinside={@}{@},
  numberstyle=\tiny\color{solarized@base01},
  showstringspaces=false,
  keywordstyle=\color{solarized@orange},
  stringstyle=\color{solarized@red}\ttfamily,
  identifierstyle=\color{solarized@blue},
  commentstyle=\CPPcommentstyle,
  directivestyle=\CPPdirectivestyle,
  emphstyle=\color{solarized@green},
  frame=single,
  rulecolor=\color{solarized@base2},
  rulesepcolor=\color{solarized@base2},
  literate={~} {\customtilde}1,
  moredelim=*[directive]\ \ \#,
  moredelim=*[directive]\ \ \ \ \#
}

\lstdefinestyle{python}
{
  language=Python,
  basicstyle=\footnotesize\ttfamily,
  basewidth={0.53em,0.44em},
  numbers=none,
  tabsize=2,
  breaklines=true,
  escapeinside={@}{@},
  showstringspaces=false,
  numberstyle=\tiny\color{solarized@base01},
  keywordstyle=\color{blue},
  stringstyle=\color{orange}\ttfamily,
  identifierstyle=\color{darkred},
  commentstyle=\color{purple},
  emphstyle=\color{green},
  frame=single,
  rulecolor=\color{solarized@base2},
  rulesepcolor=\color{solarized@base2},
  literate = {~}{\customtilde}1
             {\ as\ }{{\color{blue}\ as\ \color{black}}}3
             {.set}{{\color{black}.}{\color{darkred}set}}4
}

\lstdefinestyle{fortran}
{
  language=Fortran,
  basicstyle=\footnotesize\ttfamily,
  basewidth={0.53em,0.44em},
  numbers=none,
  tabsize=2,
  breaklines=true,
  escapeinside={@}{@},
  showstringspaces=false,
  numberstyle=\tiny\color{solarized@base01},
  keywordstyle=\color{blue},
  stringstyle=\color{orange}\ttfamily,
  identifierstyle=\color{Periwinkle},
  commentstyle=\color{purple},
  emphstyle=\color{green},
  morekeywords={and, or, true, false},
  frame=single,
  rulecolor=\color{solarized@base2},
  rulesepcolor=\color{solarized@base2},
  literate={~}{\customtilde}1
}

\lstdefinestyle{terminal}
{
  language=bash,
  basicstyle=\termplainstyle,
  numbers=none,
  tabsize=2,
  breaklines=true,
  escapeinside={@}{@},
  frame=single,
  showstringspaces=false,
  numberstyle=\tiny\color{solarized@base01},
  keywordstyle=\color{solarized@orange},
  stringstyle=\color{solarized@red}\ttfamily,
  identifierstyle=\color{black},
  commentstyle=\color{solarized@violet},
  emphstyle=\color{solarized@green},
  frame=single,
  rulecolor=\color{solarized@base2},
  rulesepcolor=\color{solarized@base2},
  morekeywords={gambit, cmake, make, mkdir, gum, python, wget, tar, cp, pippi, mpirun},
  deletekeywords={test},
  literate = {/gambit}{{/}{\color{black}}gambit}6
             {gambit/}{{\color{black}}gambit{/}}6
             {gum/}{{\color{black}}gum{/}}4
             {/include}{{/}{\color{black}}include}8
             {cmake/}{{\color{black}}cmake/}6
             {.cmake}{{.}{\color{black}}cmake}6
             {.gum}{{.}{\color{black}}gum}6
             {.tar}{{.}{\color{black}}tar}4
             {source/}{{\color{black}}source{/}}7
             { type}{{\color{black}}{}type}5
             {~}{\customtilde}1
             {math}{{\color{solarized@orange}}math}4
}

\lstdefinestyle{terminalalt}
{
  language=bash,
  basicstyle=\footnotesize\ttfamily,
  numbers=none,
  tabsize=2,
  breaklines=true,
  escapeinside={*@}{@*},
  frame=single,
  showstringspaces=false,
  numberstyle=\tiny\color{solarized@base01},
  keywordstyle=\color{solarized@orange},
  stringstyle=\color{solarized@red}\ttfamily,
  identifierstyle=\color{black},
  commentstyle=\color{solarized@violet},
  emphstyle=\color{solarized@green},
  frame=single,
  rulecolor=\color{solarized@base2},
  rulesepcolor=\color{solarized@base2},
  morekeywords={gambit, cmake, make, mkdir},
  deletekeywords={test},
  literate = {\ gambit}{{\ }{\color{black}}gambit}7
             {/gambit}{{/}{\color{black}}gambit}6
             {gambit/}{{\color{black}}gambit{/}}6
             {/include}{{/}{\color{black}}include}8
             {cmake/}{{\color{black}}cmake/}6
             {.cmake}{{.}{\color{black}}cmake}6
             {~}{\customtilde}1
}

\lstdefinestyle{text}
{
  language={},
  basicstyle=\footnotesize\ttfamily,
  identifierstyle=\color{black},
  numbers=none,
  tabsize=2,
  breaklines=true,
  escapeinside={*@}{@*},
  showstringspaces=false,
  frame=single,
  rulecolor=\color{solarized@base2},
  rulesepcolor=\color{solarized@base2},
  literate={~}{\customtilde}1
}

\lstdefinestyle{yaml}
{
  language=bash,
  escapeinside={@}{@},
  keywords={true,false,null},
  otherkeywords={},
  keywordstyle=\color{solarized@base0}\bfseries,
  basicstyle=\footnotesize\color{black}\ttfamily,
  identifierstyle=\YAMLkeystyle,
  sensitive=false,
  commentstyle=\YAMLcommentstyle,
  morecomment=[l]{\#},
  morecomment=[s]{/*}{*/},
  stringstyle=\YAMLstringstyle\ttfamily,
  moredelim=**[s][\YAMLkeystyle]{,}{:},   
  moredelim=**[l][\YAMLvaluestyle]{:},    
  morestring=[b]',
  morestring=[b]",
  literate =    {---}{{\ProcessThreeDashes}}3
                {>}{{\textcolor{solarized@red}\textgreater}}1
                {gtr}{\textgreater}1
                {grt}{\textgreater}1
                {|}{{\textcolor{solarized@red}\textbar}}1
                {\ -\ }{{\mdseries\color{black}\ -\ \negmedspace}}3
                {\}}{{{\color{black} \}}}}1
                {\{}{{{\color{black} \{}}}1
                {[}{{{\color{black} [}}}1
                {]}{{{\color{black} ]}}}1
                {~}{\customtilde}1,
  breakindent=0pt,
  breakatwhitespace,
  columns=fullflexible
}

\lstdefinestyle{gum}
{
  language=bash,
  escapeinside={@}{@},
  keywords={true,false,null,all},
  otherkeywords={},
  keywordstyle=\color{solarized@base02}\bfseries,
  basicstyle=\footnotesize\color{black}\ttfamily,
  identifierstyle=\color{solarized@magenta},
  sensitive=false,
  commentstyle=\color{solarized@cyan}\ttfamily,
  morecomment=[l]{\#},
  morecomment=[s]{/*}{*/},
  stringstyle=\footnotesize\color{solarized@base01}\mdseries\ttfamily,
  moredelim=**[l][\footnotesize\color{solarized@base02}\mdseries]{:},    
  morestring=[b]',
  morestring=[b]",
  literate =    {---}{{\ProcessThreeDashes}}3
                {grt}{{\textcolor{solarized@magenta}\textgreater}}1
                {gtr}{{\textcolor{solarized@base02}\textgreater}}1
                {/>}{{\textcolor{solarized@magenta}\textgreater}}1
                {/<}{{\textcolor{solarized@magenta}\textless}}1
                {lss}{{\textcolor{solarized@base02}\textless}}1
                {pls}{{\textcolor{solarized@magenta}+}}1
                {mns}{{\textcolor{solarized@magenta}-}}1
                {|}{{\textcolor{solarized@base02}\textbar}}1
                {\ -\ }{{\mdseries\color{solarized@base02}\ -\ \negmedspace}}3
                {\}}{{{\color{solarized@base02} \}}}}1
                {\{}{{{\color{solarized@base02} \{}}}1
                {[}{{{\color{solarized@base02} [}}}1
                {]}{{{\color{solarized@base02} ]}}}1
                {~}{\customtilde}1,
  breakindent=0pt,
  breakatwhitespace,
  columns=fullflexible
}

\lstdefinestyle{mathematica}
{
  language={Mathematica},
  basicstyle=\footnotesize\ttfamily,
  basewidth={0.53em,0.44em},
  numbers=none,
  tabsize=2,
  breaklines=true,
  postbreak=,
  escapeinside={@}{@},
  numberstyle=\tiny\color{black},
  showstringspaces=false,
  numberstyle=\tiny\color{solarized@base01},
  keywordstyle=\color{solarized@orange},
  stringstyle=\color{solarized@red}\ttfamily,
  identifierstyle=\color{solarized@orange}\ttfamily,
  commentstyle=\color{solarized@gray}\ttfamily,
  directivestyle=\color{solarized@orange}\ttfamily,
  emphstyle=\color{solarized@green},
  frame=single,
  rulecolor=\color{solarized@base2},
  rulesepcolor=\color{solarized@base2},
  literate={~} {\customtilde}1,
  moredelim=*[directive]\ \ \#,
  moredelim=*[directive]\ \ \ \ \#,
  mathescape=false
}

\newcommand{\doublecross}[2]{\hyperref[#2]{\textbf{#1}}}
\newcommand{\doublecrosssf}[2]{\hyperref[#2]{\textbf{\textsf{#1}}}}

\newcommand{\startglossary}{\section{Glossary}\label{glossary}Here we explain some terms that have specific technical definitions in \GB.\begin{description}}
\newcommand{\finishglossary}{\end{description}}

\newcommand{\bcode}{\begin{lstlisting}}
\newcommand{\ecode}{\end{lstlisting}}

\newcommand{\eV}{\ensuremath{\text{e}\mspace{-0.8mu}\text{V}}\xspace}
\newcommand{\MeV}{\text{M\eV}\xspace}
\newcommand{\GeV}{\text{G\eV}\xspace}

\newcommand{\gambit}{\textsf{GAMBIT}\xspace}

\newcommand{\darkbit}{\textsf{DarkBit}\xspace}
\newcommand{\cosmobit}{\textsf{CosmoBit}\xspace}

\newcommand{\GB}{\gambit}

\newcommand\YAML{\textsf{YAML}\xspace}

\newcommand\beq{\begin{equation}}
\newcommand\eeq{\end{equation}}

\newcommand{\dd}{\mathrm{d}}

\newcommand{\fourvec}[4]{\begin{pmatrix}#1\\#2\\#3\\#4\end{pmatrix}}
\renewcommand{\vec}[1]{\boldsymbol{#1}}
\newcommand{\updated}[1]{#1}
\newcommand{\moreupdated}[1]{#1}

\newcommand{\e}{\mathrm{e}}
\newcommand{\mpl}{M_\text{P}}
\newcommand{\mplred}{\overline{m}_\text{P}}

\newcommand{\class}{\textsf{CLASS}\xspace}
\newcommand{\CB}{\textsf{CosmoBit}\xspace}
\newcommand{\micro}{\textsf{micrOMEGAs}\xspace}
\newcommand{\plc}{\textsf{plc}\xspace}
\newcommand{\exoclass}{\textsf{ExoCLASS}\xspace}
\newcommand{\darkages}{\textsf{DarkAges}\xspace}

\newcommand{\LCDM}{$\Lambda$CDM\xspace}
\newcommand{\dnm}[2]{\newcommand{#1}{\ensuremath{#2}\xspace}}
\dnm{\TR}{T_\text{R}}
\dnm{\TD}{T_\text{dec}}
\dnm{\TCMB}{T_\text{\tiny CMB}}
\dnm{\Neff}{N_\text{eff}}
\dnm{\DNeff}{\Delta\Neff}
\dnm{\Yp}{Y_\text{p}}
\dnm{\scalef}{R}

\dnm{\ax}{a}
\dnm{\ma}{m_a}
\dnm{\gagg}{g_{a\gamma}}
\dnm{\tax}{\tau_a}

\newcommand{\sn}{SN1987A\xspace}
\newcommand{\cth}{\cos\theta}
\newcommand{\sth}{\sin\theta}
\newcommand{\cp}{\cos\phi}
\newcommand{\sinp}{\sin\phi}

\begin{document}

\title{Cosmological constraints on decaying axion-like particles: a global analysis}

\author[1]{Csaba Bal\'azs,}
\author[2]{Sanjay Bloor,}
\author[3,4]{Tom\'as E.\ Gonzalo,}
\author[5,6]{Will Handley,}
\author[4,7]{Sebastian Hoof,}
\author[3,4]{Felix Kahlhoefer,}
\author[1,8]{Marie Lecroq,}
\author[9]{David J.~E.\ Marsh,}
\author[2,10,11]{Janina J.\ Renk,}
\author[2,11]{Pat Scott}
\author[3,12]{and Patrick St\"ocker}

\emailAdd{tomas.gonzalo@kit.edu, hoof@kit.edu, kahlhoefer@kit.edu, stoecker@physik.rwth-aachen.de}

\affiliation[1]{\monash}
\affiliation[2]{\imperial}
\affiliation[3]{\aachen}
\affiliation[4]{\kitTTP}
\affiliation[5]{\cambridge}
\affiliation[6]{\kicc}
\affiliation[7]{\gottingen}
\affiliation[8]{\iap}
\affiliation[9]{\kings}
\affiliation[10]{\okc}
\affiliation[11]{\uq}
\affiliation[12]{\kitIAP}

\date{Received: date / Accepted: date}

\abstract{
Axion-like particles (ALPs) decaying into photons are known to affect a wide range of astrophysical and cosmological observables.
In this study we focus on ALPs with masses in the keV--MeV range and lifetimes between $10^4$ and $10^{13}$ seconds, corresponding to decays between the end of Big Bang Nucleosynthesis and the formation of the Cosmic Microwave Background (CMB).
Using the \cosmobit module of the global fitting framework \GB, we combine state-of-the-art calculations of the irreducible ALP freeze-in abundance, primordial element abundances (including photodisintegration through ALP decays), CMB spectral distortions and anisotropies, and constraints from supernovae and stellar cooling.
This approach makes it possible for the first time to perform a global analysis of the ALP parameter space while varying the parameters of \LCDM as well as several nuisance parameters.
We find a lower bound on the ALP mass of around $m_a > 300\,\text{keV}$, which can only be evaded if ALPs are stable on cosmological timescales.
Future observations of CMB spectral distortions with a PIXIE-like mission are expected to improve this bound by two orders of magnitude.
}

\hfill{\small gambit-physics-2022}

\hfill{\small KCL-PH-TH/2022-23}

\hfill{\small TTP22-034}

\vspace*{-2\baselineskip}

\maketitle


\section{Introduction}\label{sec:intro}

Cosmological observables such as the Cosmic Microwave Background (CMB) have now been measured with such precision that they not only constrain the properties of particles in the primordial thermal plasma, but also even probe out-of-equilibrium species with sub-dominant cosmological abundances. For example, unstable particles that decay around the time of recombination leave an observable imprint on the CMB temperature anisotropies, even if they change the energy density of the electron-photon plasma only at the level of one part per billion~\cite{Diamanti:2013bia,Slatyer:2016qyl,Poulin:2016anj}. Earlier decays are in turn tightly constrained by observations of CMB spectral distortions~\cite{1993_Hu,1910.04619} and the primordial element abundances produced in Big Bang Nucleosynthesis (BBN)~\cite{Poulin:2015opa,Hufnagel:2018bjp,2006.14803,Depta:2020zbh}. This raises the exciting possibility of using cosmology to search for the effects of decaying exotic particles across many different cosmological eras, ranging from one second to billions of years~\cite{Chen:2003gz}.

A particularly well-motivated class of exotic particles that may be probed in this way are axion-like particles (ALPs)~\cite{Kim:1986ax,1002.0329}, which generically arise in theories with spontaneously broken global chiral symmetries and theories with compactified extra dimensions~\cite{Cicoli:2012sz}. In analogy with the QCD axion proposed to solve the strong CP problem~\cite{1978_weinberg_axion,1978_wilczek_axion}, ALPs would have an approximate shift symmetry that explains both the smallness of the ALP mass and the weakness of its couplings in terms of the large scale of symmetry breaking. Although ALPs may in general have a wide range of different couplings~\cite{Brivio:2017ije,Bauer:2020jbp}, the most phenomenologically interesting one is the coupling to photons. Indeed, there is a vast literature on how to constrain ALP-photon couplings using cosmological~\cite{hep-ph/9503293,hep-ph/9702275,1110.2895,1501.04097}, astrophysical~\cite{1201.5902,1302.1208,1512.08108} and laboratory data~\cite{Bauer:2017ris,Dolan:2017osp}. For more details, we refer the interested reader to reviews on QCD axion and ALP cosmology~\cite{1510.07633} and experimental search strategies~\cite{1801.08127}, as well as a continuously updated online repository for limits and sensitivity forecasts~\cite{zenodo_axionlimits}.

Most studies of cosmological ALP constraints make one of two assumptions on the ALP abundance. Either ALPs are assumed to be produced by the realignment mechanism~\cite{Preskill:1982cy,Abbott:1982af,Dine:1982ah,Turner:1983he,Turner:1985si,1201.5902} in such a way that their present-day abundance is consistent with the observed abundance of dark matter~(DM) in the universe, or they are assumed to be in thermal equilibrium with the bath of Standard Model~(SM) particles at some high temperature but then decay, depleting their abundance. Both of these scenarios turn out to be very tightly constrained, such that ALPs are often believed to be excluded unless they are either very light (with a mass at or below the keV-scale) or very heavy (with a mass at or above the GeV-scale)~\cite{1110.2895}. It has only recently been pointed out that a low reheating temperature can open up otherwise ruled-out parameter space for intermediate-mass ALPs, if realignment production is inefficient and their couplings are small enough to avoid attaining thermal equilibrium~\cite{2002.08370}. In this case the dominant contribution to ALP production arises from the freeze-in mechanism at temperatures close to the reheating temperature, and the resulting abundance may be small enough to evade cosmological constraints arising from their subsequent decays~\cite{Bolz:2000fu,1110.2895}.

In the present work we perform a detailed analysis of ALPs in the keV--GeV mass range with tiny couplings to SM photons, such that they never enter into thermal equilibrium in the early universe. While these ALPs may in principle be sufficiently long-lived to survive until the late universe, we focus on ALPs that decay between BBN and recombination. In this case the ALP decays affect a wide range of observables, such as
\begin{itemize}
 \item the primordial element abundances through photodisintegration of the nuclei formed during BBN,
 \item the black-body spectrum of the CMB through the injection of electromagnetic energy into the plasma,
 \item the CMB anisotropies through modifications of the recombination history,
 \item the ratio of the photon and neutrino temperatures (encoded in the relativistic number of effective degrees of freedom) through a relative increase of the former,
 \item auxiliary astrophysical observables, such as axion emission and decays from supernova SN1987A or stellar cooling.
\end{itemize}

We make use of the recently released \cosmobit module~\cite{CosmoBit} of the \GB framework~\cite{gambit,grev} to perform a combined analysis of all these observables.
Exploring the combined likelihood function has the advantage of automatically incorporating any complementary information contained in the data. Similar analyses, such as the ones presented in Refs.~\cite{1110.2895,1501.04097}, investigated the individual likelihoods instead. Compared to Ref.~\cite{1110.2895} we include additional constraints from SN1987A and a more rigorous treatment of spectral distortions and \updated{other} CMB constraints. In contrast to our work, Ref.~\cite{1501.04097} focuses on models with lifetimes $\tau_a \lesssim \SI{e7}{\s}$ and \updated{not all possible types of spectral distortions}. In addition, our analysis improves upon the results of Ref.~\cite{2002.08370} by performing a more accurate calculation of the ALP abundance as a function of the reheating temperature and by including state-of-the-art likelihoods for all cosmological data sets -- in particular CMB spectral distortions and anisotropies.

\begin{figure}
  \centering
  \includegraphics[width=5in]{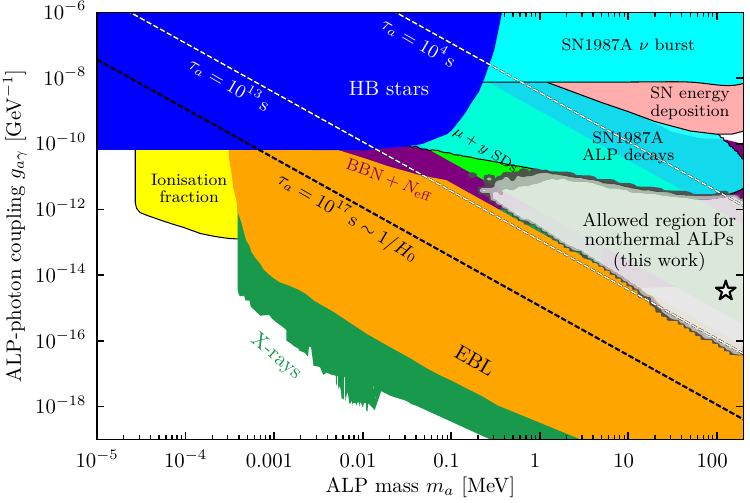}
  \caption{Summary plot of our results in the $\ma$--$\gagg$ plane. We show a part of the cosmological ALP region commonly excluded by Horizontal Branch~(HB) stars from cluster counts~\cite{1406.6053}, supernova SN1987A (see Ref.~\cite{2109.03244} and \cref{sec:SN1987A}), \updated{supernova energy deposition~\cite{2201.09890},} ionisation fraction constraints~\cite{1110.2895}, BBN and \Neff~\cite{2002.08370}, spectral distortions~\cite{1110.2895}, comparison to the extragalactic background light~(EBL)~\cite{1110.2895}, and X-ray searches~\cite{astro-ph/0603660, 0901.0011,1110.2895,2102.02207}~(we used data provided by Ref.~\cite{zenodo_axionlimits} for some of the contours). For nonthermal ALPs considered in this work, the grey-shaded region around the best-fitting point (star) is however not excluded.\label{fig:summary_plot}}
\end{figure}
A visual summary of our findings in the mass-coupling plane is shown in \cref{fig:summary_plot}. We identify large regions of viable parameter space where ALPs would constitute a sizeable fraction of DM in the very early universe, but decay away before recombination.
\updated{Considering regions of low reheating temperature $\TR \sim \SI{5}{\MeV}$, which is still consistent with BBN,} opens up parameter regions that are usually considered to be excluded by cosmology.
Moreover, by combining the individual observations into a global likelihood function we find that ALP decays can in fact improve the agreement between BBN and the CMB compared to \LCDM.
Intriguingly, the corresponding parameter regions may be fully probed by future missions that measure CMB spectral distortions.

The remainder of this work is structured as follows.
In \cref{sec:model} we discuss the freeze-in production and subsequent decay of ALPs with tiny couplings to photons.
The various ways in which these couplings can be constrained through observations and the corresponding likelihood functions are presented in \cref{sec:constraints}.
\cref{sec:results} contains the setup of our parameter scans and the results that we obtain. These results, along with a \gambit input file to reproduce them, and a plotting script, can be downloaded from Zenodo~\cite{Zenodo_CosmoALP}.

\section{Production and decay of axion-like particles coupled to photons}
\label{sec:model}

In this work we will focus on ALPs \ax with an effective coupling to photons given by
\begin{equation}
 \mathcal{L} = \frac{\gagg}{4} \ax F^{\mu\nu} \tilde{F}_{\mu\nu} \, ,
\end{equation}
where $F^{\mu\nu}$ is the electromagnetic~(EM) field strength tensor, $\tilde{F}^{\mu\nu} = \tfrac{1}{2} \epsilon^{\mu\nu\rho\sigma} F_{\rho\sigma}$ its dual, and the ALP-photon coupling $\gagg$ has mass dimension $-1$.

\paragraph{The ALP lifetime.} Axion-like particles are unstable and decay into two photons with a lifetime of $\tax = \Gamma_a^{-1}$, where
\begin{equation}
 \Gamma_a = \frac{\gagg^2 \, \ma^3}{64\pi} = \frac{1}{\SI{1.32e8}{\s}} \left(\frac{\gagg}{\SI{1e-12}{\GeV^{-1}}}\right)^2 \left(\frac{\ma}{\SI{10}{\MeV}}\right)^3 \, .
\end{equation}
We can broadly distinguish three different lifetime regimes:
\begin{enumerate}
 \item ALPs decay before neutrino decoupling ($\tax \lesssim \SI{1}{\s}$). In this case the decay photons will fully thermalise and not affect any cosmological observables.
 \item ALPs decay between neutrino decoupling and the present day ($\SI{1}{\s} \lesssim \tax \lesssim \SI{e17}{\s}$). In this case ALPs may significantly affect various cosmological observables.
 \item ALPs survive until the present day ($\tax \gtrsim \SI{e17}{\s}$). In this case ALPs contribute to the DM abundance; constraints in the mass range of interest arise from searches for mono-energetic X-ray or $\gamma$-ray photons.
\end{enumerate}
The second case is particularly interesting because non-trivial constraints on the ALP parameter space can be expected. Indeed, the previous studies mentioned in \cref{sec:intro} showed that there are very tight constraints on EM decays of DM particles during or after recombination, or during BBN.
Less attention has been given to the case of ALPs decaying between BBN and recombination, i.e.\ $\SI{e4}{\s} < \tax < \SI{e13}{\s}$, which will be the focus of this work.

\paragraph{The ALP abundance.} In addition to $\ma$ and $\gagg$, the two model parameters introduced above, the ALP abundance can be introduced as a third parameter. The abundance will be decisive for the subsequent analysis. Since in our set-up there are no processes that can produce ALPs in the temperature range that we will be interested in,\footnote{A possible exception would be the inverse decay $\gamma \gamma \to a$. This process however turns out to be irrelevant in the viable parameter space of the model.} the comoving number density $a^3 n_a$ only changes through ALP decays. We can therefore define the would-be present-day abundance in the absence of decays $n_{a,0}^{\tax \to \infty}$ and the corresponding would-be dark matter fraction
\begin{equation}
 \xi = \frac{\ma \, n_{a,0}^{\tax \to \infty}}{\rho_{\text{DM},0}} \, ,
\end{equation}
where $\rho_{\text{DM},0}$ is the present-day DM density. Because the comoving DM density is constant, the ALP fraction is temperature-independent in the absence of decays. When including ALP decays, however, the actual present-day abundance of ALPs becomes completely negligible for the lifetimes considered in this work. Hence, $\xi$ is not directly constrained by the observed DM abundance, and it is possible to have $\xi > 1$.

To study the cosmological effects of ALPs at the keV--MeV scale, Ref.~\cite{1110.2895} assumes that ALPs enter thermal equilibrium in the early universe and subsequently undergo thermal freeze-out. In this case, which has also been considered in Ref.~\cite{1501.04097}, $\xi$ can be calculated in terms of the other model parameters and, typically, $\xi \gg 1$.

Here we instead consider the case where ALPs' interactions are sufficiently weak that they never thermalise. This assumption implies an upper bound on the reheating temperature $\TR$: because ALPs couple to photons through an effective interaction, the production rate of ALPs for $T \gg \ma$ scales as $\Gamma_\text{prod} \sim \alpha T^3 \gagg^2$, where $\alpha \approx 1/137$ is the EM fine-structure constant. Thermalisation occurs if $\Gamma_\text{prod} > H$, where $H = \dot{\scalef}/\scalef$ is the Hubble rate in terms of the scale factor of the universe \scalef.\footnote{We denote the scale factor by \scalef instead of $a$ to avoid confusion with the axion field.} During radiation domination the Hubble rate depends only on the total energy density in radiation $\rho_{\rm rad}$:
\begin{equation}
  H^2(T) \simeq \frac{\rho_{\rm rad}}{3\mplred^2} = \frac{\pi^2}{90 \, \mplred^2} \, g_\ast(T) \, T^4 \approx \num{0.11} \, g_\ast(T) \, \frac{T^4}{\mplred^2} \, ,
  \label{eq:Hubble_rad}
\end{equation}
where $g_\ast$ denotes the number of effective relativistic degrees of freedom and $\mplred \equiv \mpl/\sqrt{8\pi} \approx \SI{2.435e18}{\GeV}$ is the reduced Planck mass. Hence, to first approximation, no thermalisation occurs if $\gagg$ satisfies
\begin{equation}
  \gagg \lesssim \SI{e-7}{\GeV^{-1}} \left(\frac{\SI{5}{\MeV}}{\TR}\right)^{\frac{1}{2}} \, . \label{eq:thermalisation_bound}
\end{equation}

If ALPs never enter into thermal equilibrium, their abundance depends on the details of how inflation and reheating proceed, and on the entire subsequent cosmological evolution. In particular, there may be a relevant contribution to the ALP abundance from the vacuum realignment mechanism, which depends on the initial misalignment angle of the ALP field, or from the decay of cosmic strings. For all practical purposes one can therefore treat the ALP fraction $\xi$ as a free parameter.

When doing so, it is however essential to realise that $\xi$ cannot be made arbitrarily small. The reason is that there is an irreducible\footnote{\moreupdated{We emphasise that this argument implicitly assumes that \gagg is constant during the epochs of interest. If the ALP is generated by breaking a $\mathrm{U}(1)$ symmetry, the associated radial modes have masses $m_r \gtrsim 1/\gagg$, and are thus very heavy. In the case of low \TR we therefore expect that these modes (and hence \gagg) do not evolve. Nevertheless, if \gagg is allowed to vary, the ALP abundance may be suppressed further. The investigation of such scenarios requires a more detailed description of reheating, involving e.g.\ an early phase of matter domination, which is beyond the scope of this work.}} contribution to the ALP abundance from the freeze-in mechanism, i.e.\ the direct production of nonthermal ALPs from the thermal bath of SM particles via the Primakoff effect~\cite{Bolz:2000fu,1110.2895}: $\gamma \ell^\pm \to \ax \ell^\pm$, where $\ell^\pm$ denotes the charged leptons.\footnote{Analogous Primakoff processes involving charged hadrons give a negligible contribution for the temperature range that we consider.} Let us therefore briefly review the freeze-in production of ALPs and calculate the corresponding contribution to $\xi$.

The freeze-in production is dominated by temperatures close to \TR, so that the final abundance of ALPs will be proportional to \TR~\cite{Bolz:2000fu,1402.7335}. The resulting bound on \TR is substantially stronger than the one obtained from the requirement of nonthermalisation via \cref{eq:thermalisation_bound}. Conversely, as \TR cannot be arbitrarily low, we obtain an irreducible contribution to the abundance of ALPs in the early universe.

Let us focus on temperatures below the muon mass $m_\mu$ for simplicity and consider the case that the ALP mass is negligible: $\ma \ll T \ll m_\mu \sim \SI{100}{\MeV}$. The Boltzmann equation governing the evolution of the ALP number density $n_a$ is then given by~\cite{Bolz:2000fu}
\begin{equation}
  \frac{\dd n_a}{\dd t}+3H n_a = \frac{\alpha \, \zeta(3) \, \gagg^2 \, T^6}{12\pi^2}\left[ 2 \ln\bigg(\frac{T}{m_\gamma}\bigg) + 0.8194 \right] \; , \label{eq:freeze-in}
\end{equation}
where $\zeta(3) \approx 1.202$ is {Ap\'ery's} constant and $m^2_\gamma = 4 \pi \alpha T^2 / 9$ is the plasmon mass for a fully ionised universe.\footnote{We emphasise that ALPs are typically produced with relativistic energy, such that it is not directly possible to infer the evolution of $\rho_a$ from eq.~\eqref{eq:freeze-in}. Only at sufficiently late times, when $T \ll m_a$, can we make use of the relation $\rho_a = m_a n_a$.}
Defining $x = \TR / T$ and $Y_a = n_a / s$ with $s = \tfrac{2\pi^2}{45} g^s_\ast(T) T^3$ and $g^s_\ast = g_\ast = 10.75$ in the MeV range, this becomes
\begin{equation}
  \frac{\dd Y_a}{\dd x} = \num{7.82e-5} \; \frac{\gagg^2 \, \mplred \, \TR}{x^2} \, .
\end{equation}
This equation can be readily integrated from $x = 1$ to $x \to \infty$, to yield
\begin{equation}
  \xi = \num{7.82e-5} \; \frac{\ma \, \gagg^2 \, \mplred \, \TR \, s_0}{\rho_{\text{DM},0}} = \num{0.022} \left(\frac{\ma}{\SI{1}{\MeV}}\right) \left(\frac{\TR}{\SI{5}{\MeV}}\right) \left(\frac{\gagg}{\SI{e-10}{\GeV^{-1}}}\right)^2 \, , \label{eq:rd_alp_approx}
\end{equation}
where $s_0 = \SI{2891.2}{\cm^{-3}}$ denotes the present-day entropy density. In particular, we find that $\xi \propto \TR$. The requirement that BBN proceeds as usual implies the lower bound $\TR \gtrsim \SI{5}{\MeV}$~\cite{deSalas:2015glj,1908.10189}, which in turn leads to a lower bound on $\xi$ as a function of $\ma$ and $\gagg$.

In the derivation of \cref{eq:rd_alp_approx} we made a few simplifications that are not necessarily valid. In particular, we treated both ALPs and electrons as massless. However, when $\ma > \TR$, ALP production via the Primakoff process becomes exponentially suppressed, and the lower bound on $\xi$ is significantly relaxed. This effect can be approximately captured by multiplying \cref{eq:rd_alp_approx} by a Boltzmann suppression factor $\e^{-\ma/\TR}$.

\begin{figure}
  \includegraphics[width=6in]{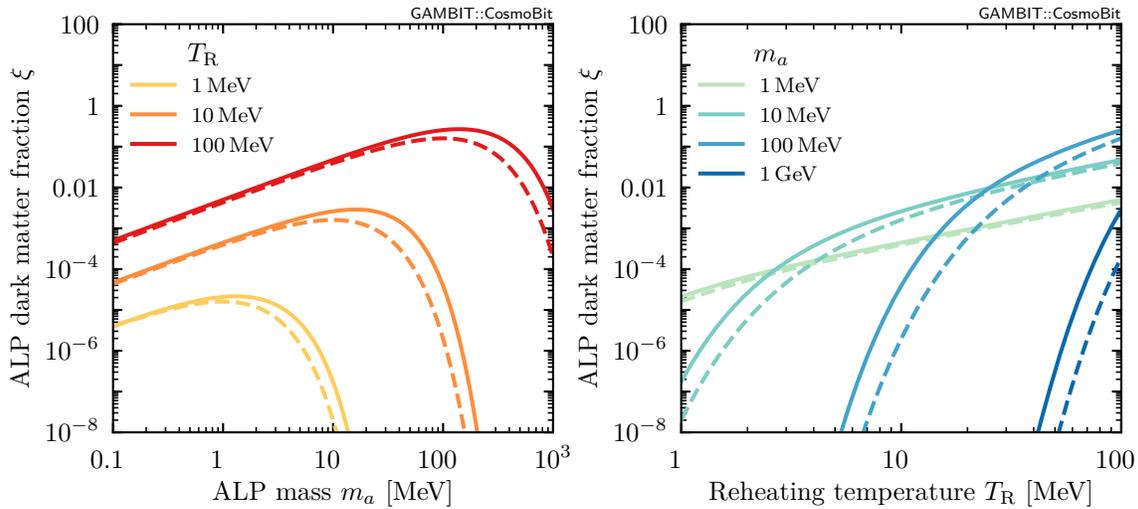}
  \caption{The irreducible freeze-in contribution to the would-be ALP dark matter fraction $\xi$ in the absence of decays as a function of $\ma$~(\textit{left}) and \TR~(\textit{right}) for a reference value of $\gagg = \SI{e-11}{\GeV^{-1}}$. We compare the numerical result from \micro~v5.2.11~(solid lines) with the approximate result from \cref{eq:rd_alp_approx} multiplied by a {na\"ive} Boltzmann suppression~(dashed lines).}\label{fig:rd_examples}
\end{figure}

To more accurately include the effect of finite masses, we also implemented the ALP model in \micro~v5.2.11. \Cref{fig:rd_examples} provides a comparison between the ALP fraction computed by multiplying \cref{eq:rd_alp_approx} by $\e^{-\ma/\TR}$, and that computed by \micro. Both approaches include quantum statistics and the plasma mass of the photon. \cref{eq:rd_alp_approx} includes a more careful treatment of the contribution of soft momenta from Ref.~\cite{Bolz:2000fu}, whereas \micro includes the sub-leading contribution from $e^+ e^- \to \ax \gamma$. Despite these differences, \cref{eq:rd_alp_approx} and the results from \micro agree very well for $m_e, \, \ma \ll \TR$. At higher masses, the curves are qualitatively similar, but diverge. In order to accurately capture the suppression of ALP production for large masses, we will always use the result from \micro in what follows. Since these calculations are computationally expensive, we perform them in advance and interpolate the tabulated results.

To conclude this section, we note that the relative contribution of ALPs to the total energy density of the universe, $\rho_a(T) / \rho_\text{tot}(T)$, will be largest shortly before they start decaying, i.e.\ for $T \approx T_\text{dec}$ with $H(T_\text{dec}) = 1/\tau_a$. Assuming that these decays happen after neutrino decoupling but before matter-radiation equality, we can estimate
\begin{equation}
 \frac{\rho_a(T_\text{dec})}{\rho_\text{tot}(T_\text{dec})} = \xi \, \frac{\rho_\text{DM}(T_\text{dec})}{\rho_\text{rad}(T_\text{dec})} = \xi \,  \frac{\Omega_\text{DM}}{\Omega_\text{rad}} \frac{T_\text{dec}}{T_0} \sim 10^{-4} \, \xi \, \left(\frac{\tau_a}{\SI{e4}{\s}}\right)^{1/2} \; .
 \label{eq:ALP_energy_ratio}
\end{equation}
We will see below that in the allowed regions of parameter space this ratio is always several orders of magnitude below unity, such that ALPs never dominate the expansion history of the universe. Nevertheless, their decays can have a number of profound effects, which we will discuss in the following section.

\section{Constraints and likelihoods}
\label{sec:constraints}

In this section we discuss the various constraints on the ALP model and the corresponding likelihood functions, which we implement within the \gambit software framework.

\subsection{CMB power spectra}
\label{sec:CMB}

We vary all parameters of the cosmological concordance model \LCDM: the present-day Hubble expansion rate $H_0$; the present-day baryon abundance $\omega_{b,0} = \rho_{b,0} h^2 / \rho_{\text{crit},0}$, where $H_0 = 100 h \, \si{\km\per\s\per\mega\parsec}$ and $\rho_{\text{crit},0} = 3 \mplred^2 H_0^2$; the present-day dark matter abundance $\omega_{\text{DM},0}$; the redshift of reionisation $z_\text{reio}$\footnote{Unlike the parametrisation used in the initial \CB release \cite{CosmoBit}, $\tau_\text{reio}$ is replaced with $z_\text{reio}$, as the latter is the quantity that is internally used within the calculations of \class and $\tau_\text{reio}$ is a derived quantity. The parametrisation used here avoids numerical instabilities which can occur when $\tau_\text{reio}$ is used as primary parameter in scenarios of strong exotic energy injection.}; the amplitude $A_\text{s}$; and the spectral index $n_\text{s}$ of the initial power spectrum. We fix the temperature of the CMB\footnote{We note that the tiny uncertainty of the CMB temperature, $\sigma(\TCMB) = \SI{5e-4}{\K}$, does become relevant in the context of CMB spectral distortions. This uncertainty is automatically marginalised over when we calculate the spectral distortion likelihood (see below).} to $\TCMB = \SI{2.72548}{\K}$, the sum of the neutrino masses to $\sum m_\nu = \SI{0.06}{\eV}$ and the pivot scale $k_0 = \SI{0.05}{\mega\parsec^{-1}}$.

For a given set of parameters we use \class~\cite{Blas:2011rf} to calculate the CMB power spectra. We then compare these predictions to data using the Planck 2018 TT,TE,EE+lowE+lensing likelihoods (referred to as  the ``baseline'' likelihood in Ref.~\cite{Aghanim:2018eyx}, see also Ref.~\cite{CosmoBit}), which are publicly available through \plc~\cite{Aghanim:2019ame}. For high multipoles we use the `lite' version of the likelihood, which requires only one additional nuisance parameter, $ A_\text{Planck} $. As done by the Planck collaboration, we apply a Gaussian prior on this parameter\cite{Aghanim:2019ame}. This is done through an additional Gaussian likelihood on this parameter, in order to allow for both, Frequentist and Bayesian analyses. The details of the interface between \gambit, \class and \plc were presented in Ref.~\cite{CosmoBit}.

Crucially, we include the effects of ALP decays on the CMB power spectra. For $\tax > \SI{e12}{\s}$, ALPs are still decaying when the electrons and protons in the primordial plasma combine to form neutral hydrogen. In this case the energy injected through these decays affects the free electron fraction and thereby delays recombination. The resulting constraints on decaying dark matter subcomponents were studied in e.g.\ Ref.~\cite{Poulin:2016anj}. The interfaces to the \class extension \exoclass, as well as to all modern versions of \class (v3.0 onwards) in \GB provide a simple way to calculate these constraints for a wide range of models, based on transfer functions calculated with \darkages~\cite{Stocker:2018avm}. The transfer functions used in this code, however, are designed to derive the impact of energy injection at redshifts around recombination and later, but not at the redshifts prior to recombination, at which spectral distortions primarily occur. At these redshifts, the universe is opaque and all injected energy will be transferred to heat up the plasma, such that it suffices to encode all impact of energy injection in a single effective efficiency factor $ f_\text{eff} $. As we assume the ALP to decay into photons exclusively, this efficiency factor is assumed to be $ f_\text{eff} = 1$. To include the constraints from energy injection in the present work we use the interface between \gambit and of \class, which was first developed in Ref.~\cite{CosmoBit}, further refined in Ref.~\cite{DMEFT}, and which we updated to work with the latest version of \class (v3.1).

For $\tax < \SI{e12}{\s}$ the dominant effect of ALP decays is a change of the photon energy density relative to those of neutrinos and baryons, which will be discussed in more detail in the following.

\subsubsection{Modifications of \Neff and $\eta_b$}\label{sec:neff}

In the absence of any beyond-the-SM physics, the only particle species that are still relativistic after BBN are photons and neutrinos. Hence the radiation energy density is given by
\begin{equation}
\rho_{\rm rad} (T_\gamma, T_\nu) = \frac{\pi^2}{15} \left( T_{\gamma}^{4} + 3 \times \frac{7}{8} T_{\nu}^{4} \right) \, ,
\label{eq:rad_dens_general}
\end{equation}
where the first term accounts for the contribution from photons and the second term accounts for the contribution from neutrinos.
As the two species are thermally decoupled, their respective temperatures $ T_\gamma $ and $ T_\nu $ evolve independently. However, because all SM processes that could modify either of the two heat baths are inefficient after neutrino decoupling, the respective temperatures are only subject to redshifting. Their ratio is therefore fixed at neutrino decoupling. In the instantaneous-decoupling approximation, the ratio's value can be derived by assuming conservation of entropy in both heat baths separately, such that $ T^\text{inst}_\nu = \left(4/11\right)^{1/3} T_\gamma $. Non-instantaneous neutrino decoupling leads to a slightly larger temperature, and it is thus convenient to define the effective number of neutrino species via~\cite{1911.04504,2005.07047,2008.01074,2012.02726}
\begin{equation}
  \Neff = \frac{3}{\big( \tfrac{4}{11}\big)^{4/3}} \left(\frac{T_\nu}{T_\gamma}\right)^4 = \num{3.0440(2)} \, ,
  \label{eq:Neff_standard}
\end{equation}
such that \cref{eq:rad_dens_general} can be rewritten as
\begin{equation}
\rho_{\rm rad} (T_\gamma) = \frac{\pi^2}{15} \left[ 1 + \frac{7}{8} \Neff \left(\frac{4}{11}\right)^{\frac{4}{3}} \right] \, T_\gamma^{4} \, .
\label{eq:rad_dens_standard}
\end{equation}

For ALPs with a lifetime $\tax > \SI{e4}{\s}$, the value of \Neff during BBN will be equal to its SM prediction, given that the contribution of ALPs produced via the freeze-in mechanism to the total energy density during radiation domination is completely negligible. Subsequent ALP decays into photons will however increase $T_\gamma$ relative to $T_\nu$ and thereby decrease the value of \Neff inferred from the CMB anisotropies. Furthermore, a modification of the photon temperature between BBN and recombination changes the baryon-to-photon ratio $\eta_b = n_b / n_\gamma$.

Including ALP decays, the evolution of the photon temperature is given by \cite{Hufnagel:2018bjp}
\begin{equation}
\frac{\dd T_\gamma}{\dd t} = \frac{15}{4\pi^2} \frac{m_a \, n_a(t) \, T_\gamma^{-3}}{\tax} - H(t) \, T_\gamma \, ,
\label{eq:dT_dt_gamma}
\end{equation}
where $ n_a(t) $ is the ALP number density, given by
\begin{equation}
n_a(t) = n_a(t_\text{ini}) \, \exp \left( -3 \int_{t_\text{ini}}^{t} \dd t' \, H(t')\right) \, \e^{-t/\tax} \, ,
\label{eq:na_t}
\end{equation}
where $t_\text{ini}$ defines the initial time, which we take to be $ 10^3\,{\rm s}$, i.e. sufficiently below the lower bound $\tau_a \geq \SI{e4}{\s}$.
The first term of \cref{eq:dT_dt_gamma} accounts for the temperature modification through the injection of entropy, whereas the second term accounts for the adiabatic cooling of the photon heat bath as the universe expands.

As $ H(t) $ depends on both temperatures $ T_\gamma $ and $ T_\nu $, it is important to also track the evolution of the neutrino temperature $ T_\nu $ as function of time. As ALPs do not interact with neutrinos, the temperature evolution is given by adiabatic cooling. Solving the corresponding differential equation yields
\begin{equation}
  T_\nu(t) = T_\nu(t_\text{ini}) \, \exp \left( -\int_{t_\text{ini}}^{t} \dd t' \, H(t')\right) \, .
\label{eq:Tnu_t}
\end{equation}
We solve the evolution of the photon temperature in \cref{eq:dT_dt_gamma} numerically up to a final time $ t_\text{fin} = \SI{2e12}{\s}$, corresponding approximately to the time of matter-radiation equality.\footnote{Note that our calculation assumes that $\tax < t_\text{fin}$. For larger ALP lifetimes, the CMB will be directly sensitive to the exotic energy injection, and the resulting constraints are much stronger than those obtained from the modification of \Neff.} In \cref{appendix:neff_impl} we provide details of our numerical implementation.

As the contribution from entropy injection is strictly positive, photons do not cool as fast as the neutrinos, and \Neff decreases between BBN and recombination.
For $ t_\text{ini} \leq t \leq t_\text{fin} $ the value of \Neff is
\begin{equation}
\Neff(t) = N_\text{eff,BBN} \left( \frac{T_\nu(t)}{T_\gamma(t)} \right)^4 \left( \frac{11}{4}\right)^{4/3},
\end{equation}
where $ N_\text{eff,BBN} \equiv \Neff(t_\text{ini}) $ is the value of \Neff at the end of BBN, given by \cref{eq:Neff_standard}. Since CMB observations are in agreement with the SM prediction, we can exclude parameter points that predict too large deviations from this value.
To calculate the resulting constraints, we pass the \LCDM parameters and the calculated value of $\Neff(t_\text{fin})$ to \class.

Furthermore, a change in $T_\gamma$ also implies a change in the comoving photon number density and hence in the baryon-to-photon ratio. As $\eta_b(t_\text{fin}) \approx \eta_{b,0}$ is tightly constrained, we invert the calculation to obtain the baryon-to-photon ratio for earlier times $ t_\text{ini} \leq t \leq t_\text{fin} $:
\begin{equation}
\eta(t) = \eta(t_\text{fin}) \left(\frac{T_\gamma(t_\text{fin})}{T_\gamma(t)}\right)^3 \exp \left( 3 \int_{t}^{t_\text{fin}} \dd t' \, H(t') \right) \, .
\label{eq:eta_ratio_t}
\end{equation}
The baryon-to-photon ratio during BBN is then simply given by $\eta_{b,\text{BBN}} = \eta_b(t_\text{ini})$.

Note that the discussion above can be easily generalised to the case of additional contributions from dark radiation during BBN, viz.\ $N_\text{eff,BBN} = 3.044 + \DNeff$.
In this case the energy density in dark radiation scales exactly as the neutrino energy density, and the ratio $N_\text{eff,BBN} / N_\text{eff,CMB}$ is independent of the initial value of $N_\text{eff,BBN}$.

\subsection{Primordial nucleosynthesis}
\label{sec:BBN}

Axion-like particles with $\tax > \SI{e4}{\s}$ decay after BBN, and the synthesis of primordial elements thus proceeds as in standard cosmology.
Still, BBN constraints do play an important role for this model, as the decays of heavy ALPs with masses above $\ma \gtrsim \SI{1}{\MeV}$ may contribute to the photodisintegration of some of the light elements~\cite{Depta:2020zbh}.
The evolution of the primordial element abundances is thus affected by the injected EM spectrum through ALP decays and can be computed by solving the Boltzmann equation~\cite{Depta:2020mhj}
\begin{equation}
 \frac{\dd Y_N}{\dd t} = \sum_j\int_0^\infty \dd E f_\gamma \left( Y_j \, \sigma_{j\gamma \to N} - Y_N \, \sigma_{N\gamma\to j} \right) \, ,
\end{equation}
where $N \in \{n, p, \ce{D}, \ce{^3H}, \ce{^3He}, \ce{^4He}, \ce{^6Li}, \ce{^7Li}, \ce{^7Be}\}$ are the primordial elements, $Y_N = n_N/n_b$ their abundances, $\sigma_p$ the cross sections for the various disintegration processes of the light elements, and $f_\gamma$ the phase space distribution of nonthermal photons produced through ALP decays.
Photons with energy $E_\gamma \gtrsim m_e^2 / (22 T)$, where $m_e$ denotes the electron mass, disappear efficiently via $e^+e^-$ creation~\cite{Kawasaki:1994sc}. Hence, disintegration processes only become efficient for temperatures $T \lesssim m_e^2 / (22 E_\text{th})$, where $E_\text{th}$ denotes the threshold energy for each nuclear reaction.
As typical threshold energies lie in the MeV range, photodisintegration becomes relevant only for $T \ll \SI{10}{\keV}$, well after the end of BBN.
The various disintegration processes and their corresponding threshold energies are given in \cref{tab:thresholds}.

\begin{table}
\centering
\caption{Threshold energies of dissociation processes considered in this work.}
\label{tab:thresholds}
{
  \hfill
  \begin{tabularx}{0.46\textwidth}{>{\collectcell{\ensuremath}}X<{\endcollectcell} S}
  \toprule
  \multicolumn{1}{l}{Reaction} & \multicolumn{1}{l}{$E_\text{th}$ [MeV]} \\
  \midrule
  \ce{\phantom{^2}D} + \gamma \to n + p & 2.22 \\
  \ce{^3H} + \gamma \to \ce{D} + n  & 6.26 \\
  \ce{^3H} + \gamma \to 2n + p  & 8.48 \\
  \ce{^3He} + \gamma \to \ce{D} + p  & 5.49 \\
  \ce{^3He} + \gamma \to n + 2p  & 7.12 \\
  \ce{^4He} + \gamma \to \ce{^3H} + p  & 19.81 \\
  \ce{^4He} + \gamma \to \ce{^3He} + n  & 20.58 \\
  \ce{^4He} + \gamma \to 2\ce{D} & 23.84 \\
  \ce{^4He} + \gamma \to \ce{D} + n + p  & 26.07 \\
  \bottomrule
 \end{tabularx}
 \hfill
 \begin{tabularx}{0.46\textwidth}{>{\collectcell{\ensuremath}}X<{\endcollectcell} S}
  \toprule
  \multicolumn{1}{l}{Reaction} & \multicolumn{1}{l}{$E_\text{th}$ [MeV]} \\
  \midrule
  \ce{^6Li} + \gamma \to \ce{^4He} + n + p  & 3.70\\
  \ce{^6Li} + \gamma \to \ce{^3H} + \ce{^3He}  & 15.79\\
  \ce{^7Li} + \gamma \to \ce{^3H} + \ce{^4He}  & 2.47\\
  \ce{^7Li} + \gamma \to \ce{^6Li} + n  & 7.25\\
  \ce{^7Li} + \gamma \to \ce{^4He} + 2n + p  & 10.95\\
  \ce{^7Be} + \gamma \to \ce{^3He} + \ce{^4He}  & 1.59\\
  \ce{^7Be} + \gamma \to \ce{^6Li} + p  & 5.61\\
  \ce{^7Be} + \gamma \to \ce{^4He}+ n + 2p & 9.30\\
  \bottomrule
  & \\
  \end{tabularx}
  \hfill
}
\end{table}

In scenarios where these depletion processes are sufficiently strong, the abundances of primordial elements will significantly differ from the predictions of vanilla \LCDM, which are known to match observations in pristine gas clouds and metal-poor stars fairly well.
Any sizeable modification of their abundances due to efficient disintegration will therefore be in considerable tension with measurements.

The element abundances are typically measured with respect to the hydrogen abundance (with the exception of $\ce{^3He}$), and we thus express the relevant element abundances as $[\ce{D}/\ce{H}]$, $[\ce{^3He}/\ce{D}]$, $\Yp = [\ce{^4He}/\ce{H}]$ and $[\ce{^7Li}/\ce{H}]$.\footnote{We do not consider the abundances of other primordial elements such as $n$, $\ce{^3H}$ and $\ce{^7Be}$ because they undergo further decay at late cosmological times and are totally depleted at the present time. We also do not consider $\ce{^6Li}$, for which no reliable measurement of the primordial abundance exists.}

For most element abundances, the most up-to-date measurements can be found in~\cite{PDG20}, with the exception of $[\ce{^3He}/\ce{D}]$, for which we have computed our own improved estimation. To estimate $[\ce{^3He}/\ce{D}] = [\ce{^3He}/\ce{H}]/[\ce{D}/\ce{H}]$, we use the analysis of Ref.~\cite{2003_Geiss}, who estimate the protosolar abundance ratios based on their measures values today. To do so, we can use their estimate for protosolar $[\ce{^3He}/\ce{H}]$, and either an estimate for protosolar $[(\ce{^3He}+\ce{D})/\ce{H}]$ or a corrected direct measurement of Jovian $[\ce{D}/\ce{H}]$. Unlike Ref.~\cite{2003_Geiss}, we use the weighted average of both methods, taking care when propagating the uncertainties. We find that $[\ce{^3He}/\ce{D}] = \num{0.82(11)}$. As the use of $[\ce{^3He}/\ce{D}]$ as a measurement is debated in the literature (see e.g.\ footnote 4 in Ref.~\cite{Depta:2019lbe}), we only use it as an upper limit by assigning an optimal likelihood to values smaller than the estimate. With this value and the recommendations from Ref.~\cite{PDG20}, the bounds that we use in this study are
\begin{align}
 \notag [\ce{D}/\ce{H}] &= \num{2.547(25)e-5} \\
 \notag [\ce{^3He}/\ce{D}] &< \num{1.03} \; \text{at 95\% confidence level~(CL)} \\
 \notag \Yp &= \num{0.245(3)} \\
 [\ce{^7Li}/\ce{H}] &= \num{1.6(3)e-10} \; .
 \label{eq:BBNabundances}
\end{align}

\subsubsection{Likelihood implementation and uncertainties}

\cosmobit uses \textsf{AlterBBN}~\cite{Arbey:2011nf,Arbey:2018zfh} to compute an initial prediction of the primordial abundances of the light elements based on a given set of cosmological parameters. It then passes these abundances to \textsf{ACROPOLIS}~\cite{Depta:2020mhj}. The latter includes the effects of photodisintegration by computing the photon spectrum in the post-BBN thermal bath and solving the Boltzmann equation for each of the primordial element abundances. Details of the \gambit interface to \textsf{ACROPOLIS} can be found in \cref{app:backends}.

Both \textsf{AlterBBN} and our interface to \textsf{ACROPOLIS} compute ratios of abundances to the hydrogen abundance $[\ce{Y/H}]$, from which we calculate $[\ce{^3He}/\ce{D}] = [\ce{^3He}/\ce{H}]/[\ce{D}/\ce{H}]$ in order to match the measured value of $[\ce{^3He}/\ce{D}]$ in \cref{eq:BBNabundances}. \cosmobit then computes a multi-dimensional Gaussian likelihood for all abundances, taking into account correlated uncertainties (see Ref.~\cite{CosmoBit} for details):
\begin{equation}
 -2\ln\mathcal{L} = f(\mathcal{P},\mathcal{O}) \, \mathcal{C}^{-1} \, f(\mathcal{P},\mathcal{O})^T + (2\pi)^n \det(\mathcal{C}),
 \label{eq:BBNloglike}
\end{equation}
where $\mathcal{P}$ is the vector of predicted abundances, $\mathcal{O}$ the vector of observed abundances, and $\mathcal{C}$ the covariance matrix. The function $f(\mathcal{P},\mathcal{O})$ is introduced to differentiate elements for which the measurement is only an upper limit, such as $\ce{^3He}$. Therefore, $f(\mathcal{P},\mathcal{O})_i = \mathcal{P}_i - \mathcal{O}_i$ for all elements except for $\ce{^3He}$, for which $f(\mathcal{P},\mathcal{O})_{\ce{^3He}} = \text{max}(\mathcal{P}_{\ce{^3He}} - \mathcal{O}_{\ce{^3He}},0)$.
This way, the likelihood becomes flat when the predicted $[\ce{^3He}/{D}]$ is smaller than the observed value, in line with our desire to implement the $[\ce{^3He}/{D}]$ measurement as an upper bound only.

We construct the covariance matrix $\mathcal{C}$ by computing the uncertainties for each of the abundances independently, and then combining them with correlation information. We estimate the uncertainties on each abundance for each parameter point by taking the larger of the distances between the mean and the high/low operational modes of \textsf{AlterBBN}, i.e.\ $\sigma_{Y_N} = \max(Y_N^{\text{high}} - Y_N^{\text{mean}},Y_N^{\text{mean}} - Y_N^{\text{low}})$. A detailed computation of the correlations is possible in \textsf{AlterBBN}, but is computationally too expensive for our purposes. Fortunately, the production of primordial elements in our model proceeds as in \LCDM. The correlations thus do not change significantly when varying the \LCDM parameters within their typical ranges. We therefore employ a constant correlation matrix through our analysis that matches the one computed by \textsf{AlterBBN} with default settings. We then propagate the covariance matrix when computing the corrections after photodisintegration with \textsf{ACROPOLIS}, using the transfer matrix provided~(see \cref{app:backends} for details). Finally, in \cref{eq:BBNloglike} we use a corrected covariance matrix, which matches the change to the observed ratio of abundances for $[\ce{^3He}/\ce{D}]$.

Besides their obvious use as predictions to compare with observations, the primordial abundances are also relevant for cosmological probes after BBN. In particular the primordial helium abundance \Yp affects the physics of recombination and thus is an input to the computation of the CMB power spectrum with \class (see \cref{sec:CMB}). As recombination happens at temperatures much lower than the energy threshold for the photodisintegration reactions for \Yp, we make sure to include these effects on the predicted value of \Yp that we use for computing the spectrum of the CMB.

In addition to the BBN likelihood, we also include a nuisance likelihood for the neutron lifetime $\tau_n$, as the reaction rates that produce the primordial elements are strongly affected by uncertainties in $\tau_n$. In this analysis we use the laboratory measurement of $\tau_n$ from ultra-cold neutron (``bottle'') experiments, $\tau_{n,\text{bottle}} = \SI{879.4(6)}{\s}$, due to their smaller uncertainties. Effects of the uncertainty on $\tau_n$ on constraints from BBN, the CMB and other measurements were studied in Ref.~\cite{CosmoBit}.

\subsubsection{The $\ce{^7Li}$ problem}

\label{sec:lithium}

The BBN predictions for the abundance of $\ce{^7Li}$ are known to be in disagreement with observations~\cite{Fields:2011zzb}. This is usually attributed to either systematic uncertainties in the determination of the primordial lithium abundance from astrophysical systems or uncertainties in the cross sections of nuclear reactions used in the theoretical calculations. Measurements of $\ce{^7Li}$ are therefore generally not considered robust enough to be used for fits of cosmological data, and we do not include them in our scans unless explicitly stated otherwise.

Nevertheless, it is still interesting to consider physics beyond the SM as a possible solution of the $\ce{^7Li}$ problem. For example, Ref.~\cite{Depta:2020zbh} suggests that the tension between predictions and observations can be resolved via ALP decays. The idea in this work is to consider ALPs in the mass range $\SI{3.17}{\MeV} < \ma < \SI{4.44}{\MeV}$ such that the photons produced in their decay have enough energy for the photodisintegration of \ce{^7Be}, but not enough for deuterium. We will return to this discussion in \cref{sec:results}.

\subsection{Spectral distortions}\label{sec:sd}

While the CMB radiation closely follows a black-body spectrum with characteristic temperature $\TCMB \approx \SI{2.725}{\K}$~\cite{Fixsen1996,Fixsen2009}, spectral distortions~(SDs) exist due to interactions taking place during the \emph{entire} thermal history of the universe~\cite{Zeldovich:1969ff,Sunyaev:1970er,1972_Zeldovich,1975_Illarionov,1982_Danese,1991_Burigana,1993_Hu,1109.6552}. As the relevance of the underlying physical processes that induce SDs changes over time, their redshift dependence has to be taken into account. This means that SDs are -- at least in theory -- sensitive to any known or new physics that alters the energy budget of the expanding universe, thus making SDs a useful, complementary probe of e.g.\ inflation, dark matter, or new particles (see Refs.~\cite{1903.04218,1910.04619} for recent reviews).

Spectral distortions affect the photon spectrum $f(x,z)$, where $x = \omega/\TCMB$ is the dimensionless frequency as observed today and $z$ is the redshift. This can be expressed as a deviation from the perfect black-body shape, i.e.
\begin{equation}
  f(x,z) = \frac{1}{\e^x + 1} + \Delta f(x,z) \, . \label{eq:bb}
\end{equation}
Historically, SDs have been classified into different types, such as $y$-type distortions~(from the Kompaneets or Sunyaev--Zeldovich parameter~$y$) or chemical potential-type $\mu$~distortions.
However, it is sufficient to specify the heating rates of the primordial plasma as a function of redshift, and integrate their spectral intensity $\Delta I \equiv (x \TCMB)^3 \Delta f(x,z) / 2 \pi^2$ to compare to observations today.
Given the frequency-integrated heating rate~$\dot{Q}$ from the Boltzmann equation, we have~(see Ref.~\cite{1910.04619})
\begin{equation}
  \Delta I \propto \frac{\Delta \rho_\gamma}{\rho_\gamma} = \int_{0}^{\infty} \! \dd z \; \frac{\dot{Q}}{(1+z)H\rho_\gamma} \, . \label{eq:change_in_rho_gamma}
\end{equation}

\subsubsection{\updated{ALP-induced spectral distortions}}

Axion-like particles can induce SDs through ALP-photon conversion in the (known) Galactic magnetic field or a putative large-scale magnetic field of primordial origin~\cite{1988_Yanagida,0905.4865,1306.6518,1308.0314,1312.3558,1801.09701,1811.11177,1908.07534}, or through ALP decays~\cite{hep-ph/9702275,1011.3694,1110.2895}. The former process is only relevant for ALPs with $\ma \ll \si{\eV}$. Since those ALPs are much lighter than the ones considered in this work, we only focus on the effect of decaying ALPs.

The effective energy injection rate of decaying ALPs, which contributes to the heating rate in \cref{eq:change_in_rho_gamma}, is given by~\cite{1910.04619}
\begin{equation}
  \frac{\dd Q_\text{inj,$a$}}{\dd t} = f_\text{eff} \, \xi \rho_\text{DM} \, \Gamma_a \, \e^{-\Gamma_a t} \, ,
\end{equation}
where we set the efficiency factor $f_\text{eff} = 1$. This is a good assumption for decays before recombination~\cite{1910.04619}. In particular, $f_\text{eff} = 1$ is justified because the energy injection happens due to EM interactions, and we use the ``on-the-spot approximation'' of instantaneous energy injection at the time of the decay.

\begin{figure}
  \centering
  \includegraphics[width=6in]{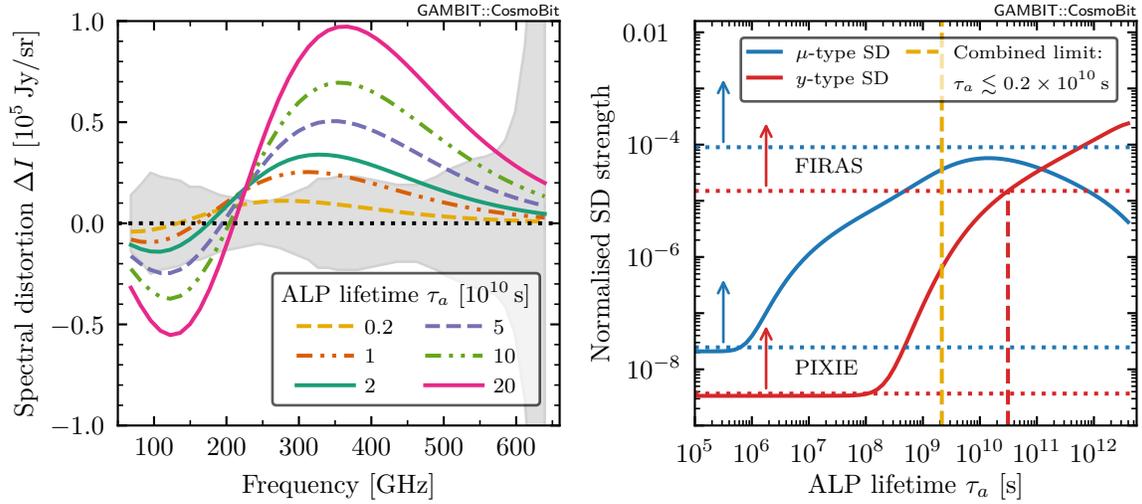}
  \caption{\textit{Left:} Predicted SDs~(coloured lines) vs uncertainties of FIRAS observations~(grey-shaded region). \textit{Right:} Predicted $\mu$ and $y$ distortions for ALPs with $\ma = \SI{100}{\MeV}$, $\xi = \num{5e-4}$, and $\TR = \SI{5}{\MeV}$ as a function of $\tau_a$. Horizontal dotted lines indicate the COBE/FIRAS limits and PIXIE forecasted limits; SDs above these lines are excluded at 95\% CL. The vertical dashed lines indicate the limits from $y$-type SDs~(red) and combined amount of SDs~(gold), respectively.\label{fig:sd_examples}}
\end{figure}
The ALP-induced SDs can in theory be calculated entirely via the unified approach of a full integration over the cosmological history outlined in \cref{eq:change_in_rho_gamma}. However, this is in practise a very expensive computation. The authors of Ref.~\cite{1910.04619} hence implemented a number of approximation schemes to simplify the calculation, which we briefly summarise in the next section. Examples for the ALP-induced spectral distortion signal in terms of $\Delta I$ are shown in the left panel of \cref{fig:sd_examples}.
There we choose ALP parameters that illustrate the power of using the general approach adopted in our analyses.

However, it is still useful to understand how ALPs can induce SDs by altering the single~\cite{1957_Kompaneets} and double~\cite{1981_Lightman} Compton rates.
Compton scattering is efficient at redshifts $z \gtrsim \num{5e4}$, where $\mu$-type distortions are sourced\updated{.
The name is due to the fact that these SDs introduce a small effective chemical potential term, often denoted by $\mu$, in the blackbody spectrum of \cref{eq:bb}.
Compton scattering then} becomes inefficient at redshifts $z \lesssim \num{5e4}$, leading to $y$-type distortions~\cite{1909.01593}. \updated{These arise as Compton scattering becomes too slow to maintain a thermal equilibrium, leading to photon energy distributions different from the $\mu$-type SDs discussed before.}

Redshifts of $z \sim \num{5e4}$ correspond to ALP lifetimes of $\tau_a \sim \SI{e10}{\s}$, indicating that the type of ALP-induced SDs can changes in different parts of the parameter space.
Spectral distortions of a given type can thus only constrain the regions of parameter space where the underlying physical interactions are relevant.

If they apply, the \updated{magnitudes of} ALP-induced $\mu$~\cite{1993_Hu} and $y$~distortions~\cite{hep-ph/9702275}~(see also Refs.~\cite{1110.2895,1312.6030,1910.04619}) can be approximated as
\begin{align}
  \mu \approx \frac{1}{\kappa_\mu} \left.\left(\frac{3\rho_a}{\rho_\gamma}-\frac{8n_a}{n_\gamma}\right)\right|_{T=\TD} \, , \quad y \approx \frac{1}{4} \left.\ln\left(\frac{\rho_a}{\rho_\gamma} + 1\right)\right|_{T=\TD} \, , \label{eq:alp_sds}
\end{align}
with $\kappa_\mu \equiv 4\zeta(2)/\zeta(3) - 3\zeta(3)/\zeta(4) \approx 2.14$ and assuming that all ALPs decay at a temperature $H(\TD) \sim \Gamma_a$.

If ALPs decay too fast or are too massive, they become inefficient in sourcing SDs, reducing the SDs in \cref{eq:alp_sds}. The more accurately predicted ALP-induced $\mu$ and $y$ distortions are shown in the right panel of \cref{fig:sd_examples}. They should be compared to the constraints from COBE/FIRAS, which are $|\mu| < \num{0.9e-4}$ and $|y| < \num{1.5e-5}$ and at 95\% CL~\cite{Mather1994,Fixsen1996,Fixsen1998}.
We also included the expected limits for a null detection from a PIXIE-like experiment~\cite[Table~1]{1910.04619}.

While $\mu$-type SDs typically set stronger constraints, this need not always be the case, as shown in \cref{fig:sd_examples}. Here we chose ALP parameters where only $y$-type SDs result in a limit of $\tau_a < \SI{3.1e10}{\s}$~(vertical dashed red line).
However, by deriving constraints using the total distortion $\Delta I$ (shown in the left panel of \cref{fig:sd_examples}), we can even deduce a stronger limit of $\tau_a < \SI{2.1e9}{\s}$~(vertical dashed gold line). This illustrates the power of the spectral likelihood approach presented in the next section.

\subsubsection{Likelihood implementation}

The authors of Ref.~\cite{1910.04619} implemented the theoretical prediction of SDs as a part of \class v3.0~\cite{Blas:2011rf}, incorporating adjusted versions of \exoclass~\cite{Stocker:2018avm}, \textsf{CosmoTherm}~\cite{1109.6552,1304.6120}, and \textsf{SZpack}~\cite{1205.5778,1211.3206}. They also provide a template likelihood for spectral distortions as a part of \textsf{MontePython}~\cite{Audren:2012wb,brinckmann2018montepython}, with specific realisations for COBE/FIRAS~\cite{Fixsen1996} data and PIXIE~\cite{1105.2044} forecasts.\footnote{The PIXIE likelihood is based on the instrument's expected sensitivity and sets the data equal to the expected signal without any new, exotic physics~(a null detection). For FIRAS, the authors of Ref.~\cite{1910.04619} use the observed sensitivity but effectively also assumes a null detection. We expect this approximation to be reasonable as the observed residuals are small compared to the uncertainties shown in the left panel of \cref{fig:sd_examples}.}

We pass the default options to \textsf{MontePython} for the SD likelihoods; in particular, this means that we use the \texttt{`exact'} scheme for the SD branching approximation method.
The \texttt{`exact'} approximation scheme is based on the Green's function method and a principle component analysis, which requires frequency binning~\cite{1304.6120,1306.5751}.

After specifying a frequency binning for each experiment, the likelihood for the observed distortions of photon spectral intensities $\Delta I_\text{obs}(\omega_i)$ at frequencies $\omega_i$ in the $i$th bin is
\begin{equation}
  \ln \mathcal{L} = -\frac{1}{2} \sum_i \left(\frac{\Delta I_\text{obs}(\omega_i) - \Delta I_\text{pred}(\omega_i)}{\delta I(\omega_i)}\right)^2 \, ,
\end{equation}
where the $\Delta I_\text{pred}(\omega_i)$ are the theory predictions by \class and the $\delta I(\omega_i)$ are the expected sensitivities, which are the sum in quadrature of the experimental noise level and theory uncertainties.
This includes foreground maps, which were accounted for in Ref.~\cite{2010.07814} and largely based on Ref.~\cite{Abitbol:2017vwa}.

One of the strengths of implementing the likelihood with the computational framework described above is that we avoid the complications of using \cref{eq:alp_sds} for $\mu$ and $y$ distortions separately. Instead, we constrain all types of SDs simultaneously, using essentially all spectral information contained in the data. Our results should thus supersede previous works such as Refs.~\cite{1011.3694,1110.2895}.

\subsection{ALP decays after SN1987A}
\label{sec:SN1987A}

The authors of Ref.~\cite{1702.02964} set limits on \ma and \gagg from ALPs emitted from core-collapse supernova \sn, using data~\cite{Chupp:1989kx} of the gamma-ray spectrometer onboard the Solar Maximum Mission (SMM)~\cite{1980_Forrest}.
The (instantaneous) emission spectrum of ALPs with energy $E_a$ is
\begin{equation}
  \frac{\dd N_a}{\dd E_a} = C \; \frac{E_a^2}{\e^{E_a/T_\text{\tiny SN}} - 1} \; \sigma_\text{P}(\ma,\,\gagg,\,E_a) \, , \label{eq:alp_spectrum_sn1987a}
\end{equation}
with $T_\text{\tiny SN} = \SI{30.6}{\MeV}$~\cite{1702.02964}. Here $C = \SI{2.54e77}{\MeV^{-1}}$ is a fit to the results of Ref.~\cite{1410.3747}, and $\sigma_\text{P}$ is the Primakoff cross section \cite{1110.2895}, for which we use a Debye screening scale of $k_\text{s} = \SI{16.8}{\MeV}$~\cite{1702.02964}

The limits arise because the emitted ALPs can subsequently decay into two photons at some distance $L_a$, following an exponential distribution with mean distance
\begin{equation}
    \ell = \frac{\beta\gamma}{\Gamma_a} = \frac{64\pi}{\gagg^2\ma^3} \, \sqrt{\frac{E_a^2}{\ma^2} - 1} \, , \label{eq:alp_decay_length}
\end{equation}
where $\beta$ and $\gamma$ are the dimensionless ALP velocity and Lorentz boost factor, respectively.

\begin{figure}
  \centering
  {
    \begin{tikzpicture}[scale=1,rotate=0]
      \input{figures/sn1987a_decay}
    \end{tikzpicture}
    \hfill
    \includegraphics[width=3.05in]{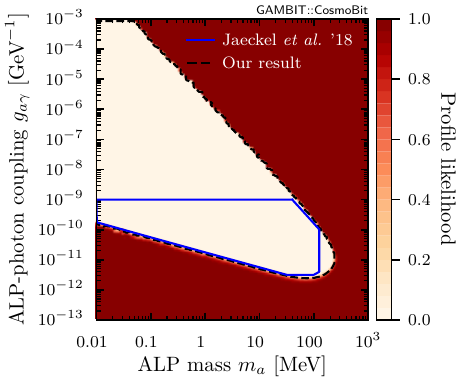}
  }
  \caption{\textit{Left:} Geometry for \sn ALPs decaying into two photons, emitted at angles $\alpha_{1/2}$ at a distance $L_a$ in the reference frame of the SMM satellite. Image credit for the SSM satellite to G.~Nelson/NASA~(JSC image library; public domain). \textit{Right:} Our profile likelihood and exclusion region of $\ma$ and $\gagg$ (dashed black line) compared to that of Ref.~\cite{1702.02964}~(blue line) at $3\sigma$ CL, assuming 1 degree of freedom.}
  \label{fig:illustration}
\end{figure}

To calculate the expected photon fluence from ALP decays, we estimate the ``success rate'', i.e.\ the fraction of photons arriving at the SMM satellite, via Monte Carlo simulations. The expected fluence is then calculated by integrating Eq.~\eqref{eq:alp_spectrum_sn1987a}, dividing by the surface of the sphere centred on the supernova with the distance to the detector as the radius ($d_\text{\tiny SN} = \SI{51.4}{\kilo\parsec}$), and multiplying by the success rate. An illustration of the setup is provided in the left panel of \cref{fig:illustration}.

Following Ref.~\cite{1702.02964}, we use a Gaussian likelihood function assuming a null result where the standard error is derived from the observed photon fluence in the energy range of \SIrange{25}{100}{\MeV}. The resulting exclusion limit is shown in the right panel of \cref{fig:illustration}. We provide details of our implementation in Appendix~\ref{appendix:sn1987a}, as well as a table of the signal prediction that ships as part of the latest version of \gambit, and publicly available Python scripts for our Monte Carlo procedure.\footnote{Available on \textsf{Github} at \url{https://github.com/marie-lecroq/ALP-fluence-calculation}\moreupdated{, tagged \texttt{v1.0}}.}

\subsection{Stellar evolution and cluster counts}

Due to their weak interactions with matter, ALPs can carry away energy from stellar interiors, affecting the evolution of e.g.\ Horizontal Branch~(HB) stars by reducing their lifetimes~\cite{Sato:1975vy,Raffelt:1990yz,book_raffelt_laboratories}. This allows one to place a limit on $\gagg$ by comparing the number of HB~stars $N_\text{HB}$ to the number of stars on the Red~Giant Branch~(RGB) $N_\text{RGB}$ in globular clusters (GCs). The associated ``$R$~parameter'' observable $R_\text{GC} \equiv N_\text{HB}/N_\text{RGB}$~\cite{1983A&A...128...94B}, is directly connected to the lifetimes and energy loss rates of these systems.

The theoretically expected $R$~parameter can be estimated from stellar evolution simulations that include the energy loss from ALPs~\cite{1512.08108,1983A&A...128...94B,Raffelt:1989xu,1406.6053},
\begin{equation}
  R_\text{GC} \approx 0.022 - 0.443 \, (1 + 0.965 \, |\gagg|)^{1/2} + 7.331 \, Y_\text{GC} \, , \label{eq:Rparameter}
\end{equation}
where $Y_\text{GC} = 0.2515(17)$~\cite{1503.08146} is an estimate for the Helium abundance in GCs. The observed value of the $R$~parameter, $R_\text{GC,obs} = \num{1.39(3)}$~\cite{1406.6053}, is consistent with the axion-free value from~(\ref{eq:Rparameter}), which is $R_\text{GC} = \num{1.42(1)}$.

\Cref{eq:Rparameter} is only valid for ALP masses much smaller than the typical core temperature of an HB~star, $\ma \ll T_\text{HB} \sim \SI{10}{\keV}$. For greater ALP masses, the Primakoff interaction cross section, which leads to the energy loss in HB stars, should be multiplied by a correction factor $F(\ma)$. Using Eq.~(A.5) of Ref.~\cite{1110.2895}, we obtain $F(\ma)$ normalised such that $F(\ma \ll T_\text{HB}) \simeq 1$. As the Primakoff cross section and the energy loss are proportional to $\gagg^2$, it is sufficient to replace $\gagg \mapsto \gagg \times \sqrt{F(\ma)}$ in \cref{eq:Rparameter} in order to apply the correction.

The $R$~parameter likelihood is part of the \darkbit \cite{darkbit} module of \gambit. It was first introduced in Ref.~\cite{Axions}, and the correction for larger masses was added in Ref.~\cite{2007.05517}. Also note that slightly stronger limits in the region around $\ma \sim \SI{100}{\keV}$ may be obtained by considering photon coalescence~\cite{2004.08399}. As the $R$~parameter constraints only marginally affect our results, we do not include these slightly improved constraints.

\subsection{Other likelihoods}

Finally, we make use of \textsf{MontePython} v3.5 to include the BOSS DR12~\cite{Alam:2016hwk} likelihood for Baryon Accoustic Oscillations and the Pantheon~\cite{Scolnic:2017caz} likelihood for type Ia supernovae. The latter likelihood requires the introduction of an additional nuisance parameter $M$ corresponding to the absolute magnitude
of a type Ia supernova, see Ref.~\cite{CosmoBit} for details.

\section{Results}
\label{sec:results}

By combining all of the likelihoods introduced in \cref{sec:constraints}, we are now in the position to perform a global scan of ALP decays in the early universe. We first present the results from a frequentist analysis  of the parameter space, which enables us to identify particularly interesting parameter regions, and then perform a Bayesian analysis, which provides complementary information on the preference between the ALP model and \LCDM.

\subsection{Frequentist scans}

The parameters included in our scans are summarised in \cref{tab:parameters}.
We emphasise that the parameter ranges for the ALP model are somewhat arbitrary and are chosen in such a way as to capture all relevant features.\footnote{We note that for the smallest ALP masses and the shortest lifetimes that we consider, the assumption of non-relativistic ALP decays is not a good approximation for ALPs produced via freeze-in, because it underestimates the energy density of ALPs. Nevertheless, we find that this parameter region is robustly excluded in our scans. A more accurate treatment would only strengthen this exclusion and is therefore unnecessary.
}
In other words, while viable parameter regions exists also beyond the scan ranges, they do not lead to any qualitative changes in our conclusions.
The scan ranges for the \LCDM and other nuisance parameters are chosen so that they fully encompass the allowed parameter regions.
\updated{
Note that this statement is not in contradiction with our treatment of the reheating temperature, which we fix to $\TR = \SI{5}{\MeV}$.
This choice is conservative since limits that depend on \TR can typically be evaded in the region of low \TR.
Thanks to the various parameter degeneracies of the ALP abundance -- schematically captured in \cref{eq:rd_alp_approx} -- and our choice of leaving $\xi$ as a free parameter, we can still access the parameter space studied in previous works such as Ref.~\cite{2002.08370}.
In terms of our numerical results, our choice of \TR only has a mild effect: for example, going from $\TR = \SI{5}{\MeV}$ to $\TR = \SI{100}{\MeV}$ will change the lowest allowed ALP mass (at 95\% CL) from $\ma \lesssim \SI{0.25}{\MeV}$ to $\ma \lesssim \SI{0.71}{\MeV}$.
}

\begin{table*}[t]
  \centering
  \caption{List of model and nuisance parameters included in our scans and their corresponding scan ranges.}
  \label{tab:parameters}
  \begin{tabular}{lcrl}
    \toprule
    \multicolumn{2}{l}{\textbf{Model parameter}} & \multicolumn{2}{l}{\textbf{Scan range}} \\
    \midrule
    ALP mass & \ma & $[0.001,\, 200]$ & \si{\MeV} \\[1mm]
    ALP lifetime & \tax & $[10^4,\, 10^{13}]$ & \si{\s} \\[1mm]
    ALP abundance & $\xi$ & $[10^{-12},\, 10^2]$ & \\[1mm] \midrule
    Baryon abundance & $\omega_b$ & $[0.020,\, 0.024]$ & \\[1mm]
    Dark matter abundance & $\omega_\text{DM}$ & $[0.10,\, 0.13]$ & \\[1mm]
    Hubble constant & $H_0$ & $[62,\, 74]$ & $\si{\km\per\s\per\mega\parsec}$ \\[1mm]
    Redshift of reionisation & $z_\text{reio}$ & $[4.5,\, 9.5]$ & \\[1mm]
    Primordial curvature & $\ln(10^{10} A_s)$ & $[2.9,\, 3.2]$ & \\[1mm]
    Scalar spectral index & $n_s$ & $[0.9,\, 1.1]$ & \\[1mm]
    \midrule
    Neutron lifetime & $\tau_n$ & $[875,\, 895]$ & \si{\s} \\[1mm]
    Planck nuisance parameter & $A_\text{Planck}$ & $[0.9,\, 1.1]$ & \\[1mm]
    Pantheon nuisance parameter & $M$ & $[-20,\, -18]$ & \\[1mm]
    \bottomrule
  \end{tabular}
\end{table*}

To perform comprehensive scans of the parameter space we employ the differential evolution sampler \textsf{Diver} v1.0.4~\cite{ScannerBit}, using a population size of 20,000 and a convergence threshold of \num{e-8}. We carried out additional scans targeted at specific regions of parameter space in order to ensure that the likelihood function is fully explored. As we perform a frequentist analysis based on the profile likelihood, we simply combine samples from different scans. The results shown below are based on a total of 33.5 million samples and took over 700,000 CPU hours on the Joliot-Curie HPC cluster at CEA (France).

In what follows, we use Wilks' theorem to determine the exclusion regions of the profile likelihood. To do so we assume that the number of degrees of freedom corresponds to the number of free parameters -- i.e.\ two degrees of freedom for the two-dimensional plots shown below. It is well known that the assumptions of Wilks' theorem often do not hold exactly, and that care should be taken to ensure its validity when claiming a discovery~\cite[e.g.][]{1911.10237}. For the purpose of setting limits w.r.t\ the best-fitting point, however, we argue that Wilks' theorem typically presents a good approximation and avoids costly Monte Carlo simulations that do not alter significantly the findings of our analysis (see e.g.\ Ref.~\cite[appendix~D]{2007.05517} for a demonstration in the context of global fits).

\begin{figure}
 \includegraphics[width=0.494\textwidth,trim={0 42 60 0},clip]{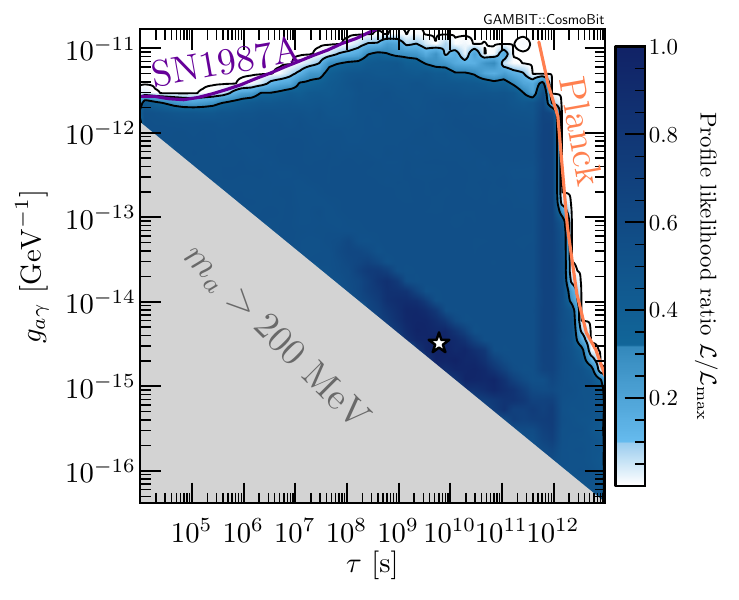}
 \includegraphics[width=0.486\textwidth,trim={65 42 0 0},clip]{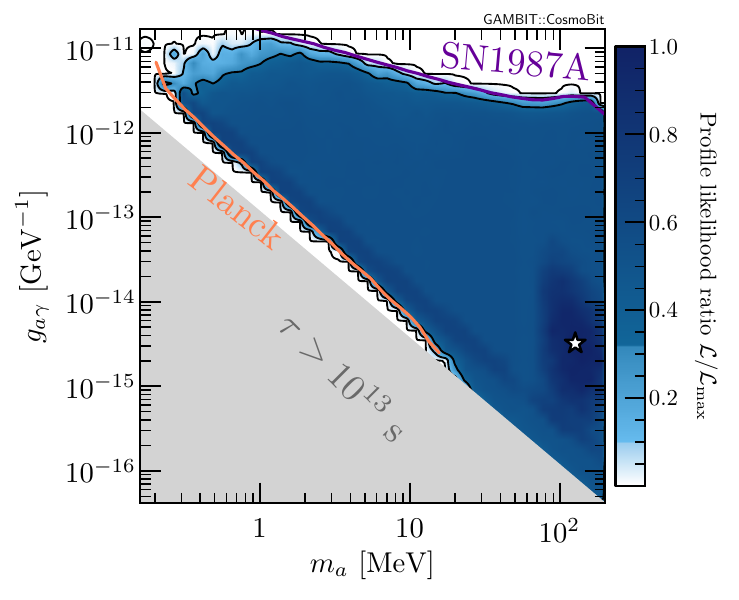}\\
 \includegraphics[width=0.494\textwidth,trim={0 0 60 0},clip]{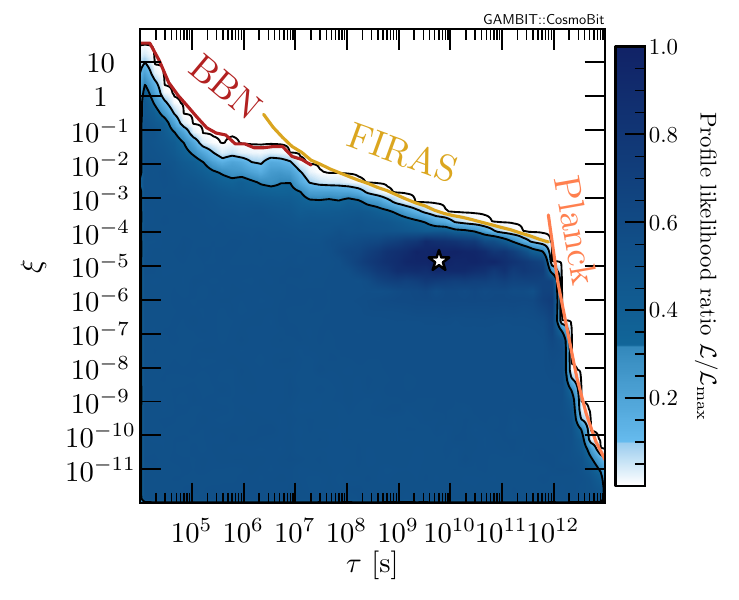}
 \includegraphics[width=0.486\textwidth,trim={65 0 0 0},clip]{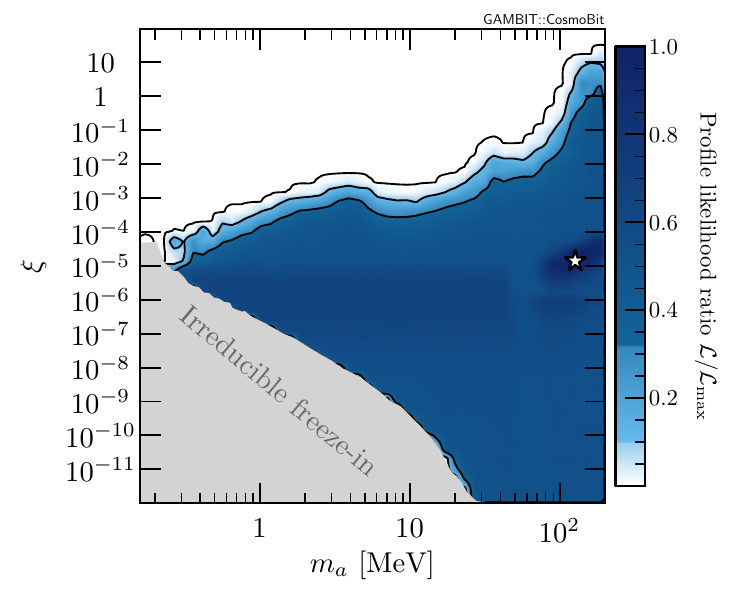}
 \caption{Profile likelihood for various combinations of the parameters. Black lines correspond to the 1$\sigma$, 2$\sigma$, and 3$\sigma$ contours around the best-fitting point (white star), whereas coloured lines are 99.7\% CL limits from the individual likelihoods. Shaded grey regions are excluded from our study as they are beyond the scan ranges.}
 \label{fig:results}
\end{figure}

Our main results are shown in \cref{fig:results}, where we plot the profile likelihood as a function of different combinations of ALP model parameters.
We remind the reader that there are three independent ALP model parameters, one of which is profiled out in each panel.
We use grey shading to indicate parameter regions that are not explored in our scans due to the chosen scan ranges for \ma and \tax.
In particular, the upper bound on \tax implies a lower bound on \gagg as a function of \ma; this, in turn, implies a lower bound on $\xi$ because of the irreducible freeze-in contribution, cf.\ \cref{eq:rd_alp_approx}.

Each of the applied constraints is sensitive to a specific combination of model parameters and can therefore be understood most naturally in one of the panels in \cref{fig:results}. For example, the astrophysical constraints are independent of $\xi$ and therefore directly constrain the combination (\ma,\,\gagg) or, equivalently, (\tax,\,\gagg). The cosmological constraints, on the other hand, are to first approximation independent of \ma. They are therefore most easily expressed in terms of ($\xi$,\,\tax). Coloured lines and labels indicate the locations of individual constraints (at 99.7\% CL) in the various panels.

We find that combining the various likelihoods excludes large parts of the ALP parameter space.
The cosmological constraints become stronger with increasing ALP lifetime and require $\xi \ll 1$, unless $\ma \gtrsim \SI{100}{\MeV}$ and $\tax \lesssim \SI{e5}{\s}$. Moreover, we find that there are no viable parameter regions for $\ma \lesssim \SI{300}{\keV}$ as a result of the \sn constraint. This is because it places an upper bound on \gagg, and therefore a lower bound on \tax, which in turn leads to strong cosmological constraints. We emphasise that this lower bound on \ma is a direct result of observational constraints and does not depend on the scan ranges that we have chosen. However, when significantly expanding the scan ranges, we find that there is a second disjoint region of allowed parameter space at smaller ALP masses, ALP lifetimes comparable to or greater than the age of the universe, and a tiny abundance. The most promising way to probe this parameter region would be through searches for X-ray lines~\cite[e.g.][]{1110.2895}, which we do not consider in the present work.

We observe that it is essential to combine different likelihoods in order to constrain the parameter space from various directions. However, we find that no relevant constraints are obtained from the modification of $N_\text{eff}$ and $\eta_b$ discussed in \cref{sec:neff}. The reason is that in the parameter regions allowed by all other constraints, the energy density of ALPs relative to radiation is simply too small to change the photon temperature in an observable way. Even for the shortest ALP lifetimes and largest ALP masses that we consider, eq.~\eqref{eq:ALP_energy_ratio} implies $\rho_a / \rho_\text{tot} < 10^{-3}$, which restricts the relative modification of $N_\text{eff}$ and $\eta_b$ to a similar level. Furthermore, we find that once all other constraints are imposed, the likelihood for $R_\text{GC}$ has no effect on the remaining parameter space, since in the allowed parameter region the ALPs are too heavy to be produced in horizontal branch stars.

\begin{figure}
 \centering
 \includegraphics[width=0.6\textwidth]{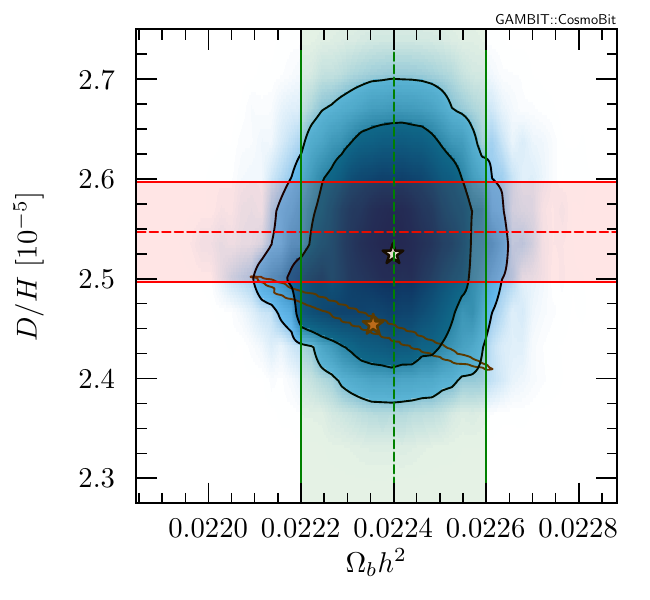}
  \includegraphics[width=0.37\textwidth]{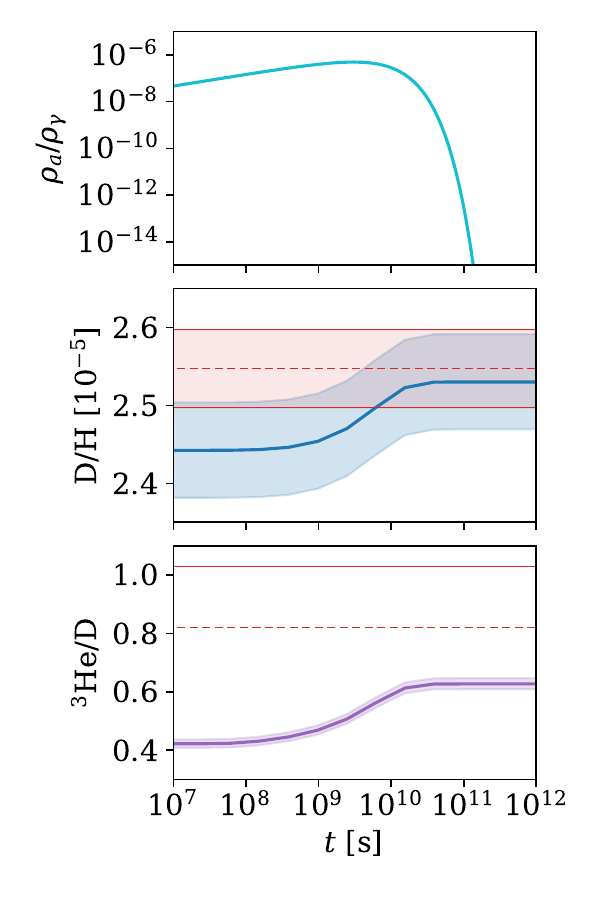}
 \caption{\emph{Left:} Profile likelihood in the plane of baryon density $\Omega_b h^2$ and deuterium abundance $\ce{D}/\ce{H}$. Black lines correspond to the 1$\sigma$ and 2$\sigma$ contours around the best-fitting point for the ALP model (white star). The brown line indicates the 2$\sigma$ contour around the best-fitting point in \LCDM~(brown star). Horizontal and vertical lines mark the central values and the 2$\sigma$ intervals of the measured $\Omega_b h^2$~\cite{Aghanim:2018eyx} and $\ce{D}/\ce{H}$ (see \cref{sec:BBN}) values, respectively. \emph{Right:} Time evolution of the ALP energy density (relative to the energy density of photons) as well as of the abundance ratios $\ce{D}/\ce{H}$ and $\ce{^3He}/\ce{D}$ for the best-fit point. Dashed (solid) red lines indicate the observed values ($2\sigma$ bounds) of the abundance ratios, the shaded bands represent theoretical uncertainties, which for the case of $\ce{D}/\ce{H}$ are comparable to the measurement uncertainties.}
 \label{fig:observables}
\end{figure}

We find that large regions of parameter space feature a relatively constant likelihood, which corresponds to the case that the ALP has negligible impact on cosmological observables. However, we also find a small region that is slightly preferred by data. The best-fitting point is found at $\ma = \SI{126.1}{\MeV}$, $\tax = \SI{6.04e9}{\s}$ (corresponding to $\gagg = \SI{3.31e-15}{\GeV^{-1}}$), and $\xi = \num{1.48e-05}$ (compatible with the irreducible freeze-in contribution for $\TR \lesssim \SI{4}{\GeV}$) and gives $-2 \Delta \ln \mathcal{L} = 1.51$ compared to the case of negligible ALP abundance.

To understand the origin of this improvement, we show in \cref{fig:observables} the allowed parameter region in terms of the deuterium abundance $\ce{D}/\ce{H}$ and the baryon abundance $\Omega_b h^2$. For comparison, we show the measured $2\sigma$ CL interval for both quantities as horizontal and vertical bands (see \cref{sec:BBN} for the experimental measurement of $\ce{D}/\ce{H}$ and Ref.~\cite{Aghanim:2018eyx} for that of $\Omega_b h^2$). On the one hand, we find that in \LCDM these two quantities are tightly correlated, and it is not possible to perfectly fit both measurements simultaneously. On the other hand, in our ALP model it is possible to increase the deuterium abundance through photodisintegration of \ce{^4He}, which clearly improves the agreement with observations.\footnote{\updated{Note that the production of deuterium through photodisintegration of \ce{^4He} dominates over the photodisintegration of deuterium itself, because \ce{^4He} is much more abundant than deuterium and hence hard photons are more likely to be absorbed by the former than the latter.}}

We remind the reader that our scans include a theoretical uncertainty for the prediction of $\ce{D}/\ce{H}$, which is calculated on the fly using \textsf{AlterBBN}. Indeed, for most points the relative theoretical uncertainty is large compared to the observational uncertainty (see also Ref.~\cite{CosmoBit}). This is why the allowed parameter region of our model extends well beyond the observed confidence interval for $\ce{D}/\ce{H}$. The best-fitting point of \LCDM is hence found to lie within the $1\sigma$ allowed parameter region of the ALP model. This finding emphasises the need for an accurate treatment of uncertainties in the calculation of BBN abundances. Indeed, neglecting the theoretical uncertainty on $\ce{D}/\ce{H}$ would significantly overestimate the preference of the ALP model over \LCDM.

\begin{figure}
 \includegraphics[width=0.494\textwidth,trim={0 42 60 0},clip]{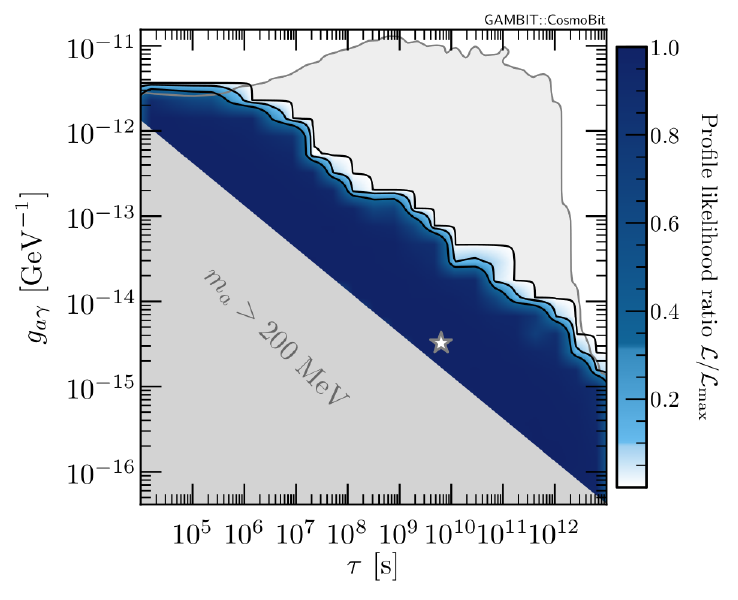}
 \includegraphics[width=0.486\textwidth,trim={65 42 0 0},clip]{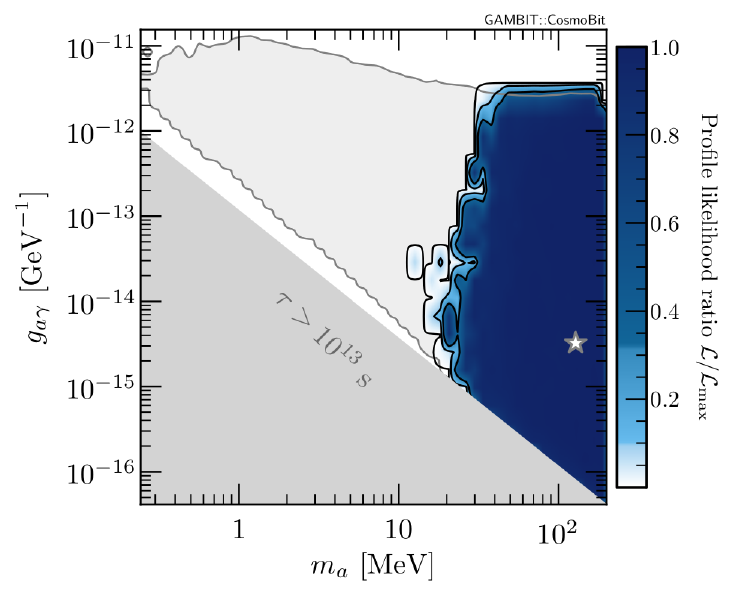}\\
 \includegraphics[width=0.494\textwidth,trim={0 0 60 0},clip]{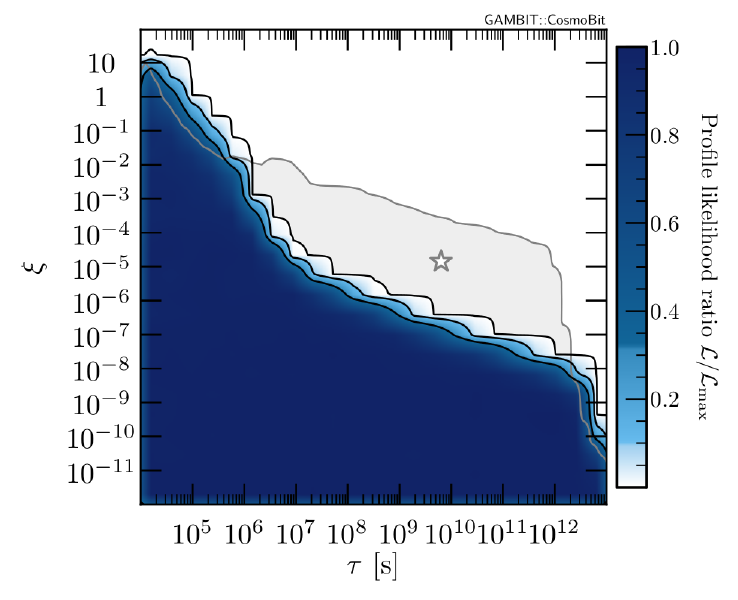}
 \includegraphics[width=0.486\textwidth,trim={65 0 0 0},clip]{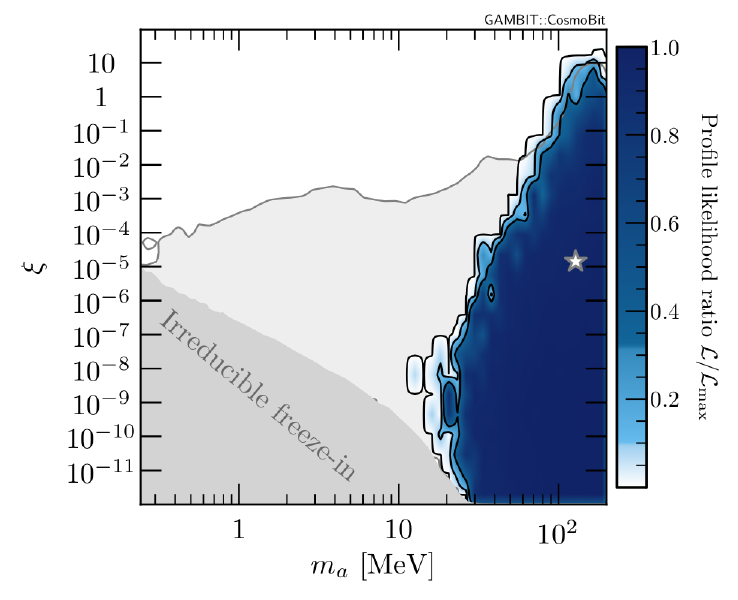}
 \caption{Profile likelihoods as in \cref{fig:results} but for the projected PIXIE limits instead of the FIRAS likelihood. Black lines again correspond to the 1$\sigma$, 2$\sigma$, and 3$\sigma$ contours around the best-fit point. The likelihood in the allowed region is very flat such that the best-fit point is not unique; it is thus not shown here. The grey shaded region corresponds to the 2$\sigma$ region around the best-fit point (grey star) of the scans with the FIRAS likelihood from \cref{fig:results}.}
 \label{fig:pixie}
\end{figure}

While the preference for our model over \LCDM is certainly not significant, it is still interesting to understand which future observations may shed further light on this issue.
We therefore performed another set of scans where we replace the COBE/FIRAS likelihood with a mock likelihood corresponding to the sensitivity of a PIXIE-like mission~\cite{1105.2044}. Such a PIXIE mock likelihood, where a null result for new physics was assumed to generate typical observed data, was first studied in Ref.~\cite{1910.04619} and made publicly available through \textsf{MontePython}.\footnote{Note that it is not currently possible in \textsf{MontePython} to combine the COBE/FIRAS likelihood and the PIXIE mock likelihood. However, since the latter is assumed to be substantially more constraining than the former, we expect results from a combination of both likelihoods to be very similar to the case with only the PIXIE mock likelihood. In any case, the PIXIE likelihood is subject to considerable unknowns at this point and should therefore only be considered an approximation.} The results are shown in \cref{fig:pixie} and are indicative of the improvements that can be expected from future measurements of CMB SDs.
We find that a null result from PIXIE would indeed significantly tighten the constraints on our model and robustly exclude the best-fit point found in the previous scan.
In other words, the best-fitting point of our main analysis including the COBE/FIRAS likelihood predicts CMB SDs that should be easily detectable with future missions, such that PIXIE would be able to determine to high significance whether photodisintegration of $\ce{^4He}$ caused by ALP decays is responsible for the large value of $\ce{D}/\ce{H}$ compared to the naive \LCDM prediction. We emphasise that future measurements of CMB anisotropies (using for example CMB-S4~\cite{CMB-S4:2016ple,Abazajian:2019eic}) would not lead to similar improvements in the constraints, since the predicted modifications of $\Delta N_\text{eff}$ and $\eta_b$ are simply too small to be observable.

Let us finally comment on the $\ce{^7Li}$ problem, \updated{which can potentially be addressed through the decays of ALPs with mass in the range $\SI{3.17}{\MeV} < \ma < \SI{4.44}{\MeV}$ (see section~\ref{sec:lithium})}. To understand if the ALP model considered in this work can solve this problem, we ran a separate scan in which the $\ce{^7Li}$ measurement is included as an additional likelihood. In agreement with the results from Ref.~\cite{Depta:2020zbh}, we find that the ALP abundances required to significantly modify the $\ce{^7Li}$ abundance must be of order $\xi \sim 1$. Such large ALP abundances are however robustly excluded by CMB SDs, which were not considered in Ref.~\cite{Depta:2020zbh}. We therefore conclude that, once all relevant constraints are taken into consideration, ALPs decaying into photons cannot resolve the $\ce{^7Li}$ problem.

To conclude, we note that we have checked explicitly that all results presented in this section are robust against additional contributions to $\Delta N_\text{eff}$ from further ultrarelativistic species. This is a direct consequence of the fact that the constraint on $N_\text{eff}$ does not dominate our exclusion limits.

\subsection{Bayesian scans}

The priors that we employ for Bayesian scans extend across the ranges quoted in \cref{tab:parameters}. For the \LCDM parameters and the nuisance parameters, all of which are relatively tightly constrained by data, we employ flat priors on each interval.\footnote{In particular, we use a flat prior for $A_\text{Planck}$ and include a Gaussian likelihood in our scans to constrain it. This procedure leads to the same posteriors as when using a Gaussian prior with no additional likelihood.} The ALP model parameters, on the other hand, can vary over many orders of magnitude. In order not to impose a specific scale, we employ logarithmic priors for these parameters.

These scans took place on the Cambridge CSD3 DiRAC cluster, with sampling by \textsf{PolyChord}~\cite{Handley:2015} v1.20.1 with \texttt{nlive}$=1000$, \texttt{nprior}$=40000$, \texttt{tol}$=10^{-10}$ and \texttt{logzero}$=-10^{101}$. These settings ensure good sampling of the prior, the likelihood peak and the broader posterior, and give control over the points thrown away at the prior level.

We performed a \LCDM run and an ALP run with the same priors on the cosmological parameters. The resulting posteriors from the two runs for the baseline cosmological parameters are indistinguishable and no correlations are found between the \LCDM parameters and the ALP parameters.
For the ALP run, we consider two effective priors on the ALP parameters: the default ``box prior'' given by the ranges in \cref{tab:parameters}, and a ``Primakoff prior'', which results from throwing away points within the box prior where the irreducible ALP abundance from the Primakoff process (for $\TR = \SI{5}{\MeV}$) is incompatible with the chosen value of $\xi$. We emphasise that in the case of the ``box prior'' we still impose the lower bound on $\xi$ at the level of the likelihood, so that the two approaches lead to indistinguishable posteriors.

\begin{figure}
  \centering
  \includegraphics[width=6in]{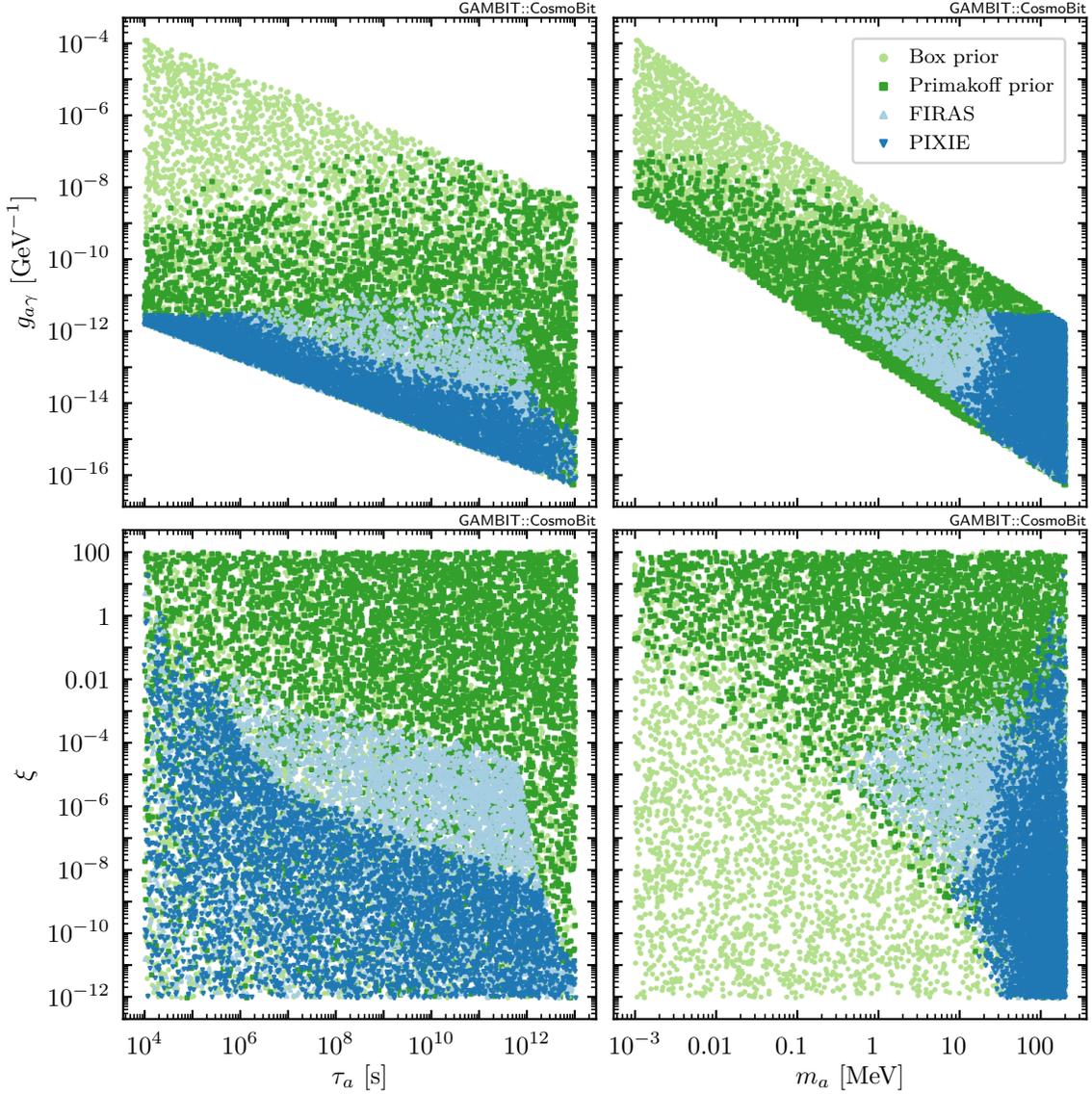}
    \caption{Bayesian scan analogue of \cref{fig:results}. \updated{Light green points represent equally-weighted samples drawn from the default box prior, dark green points are samples drawn from the more restrictive Primakoff prior (see text for details), and light and dark blue points} are samples drawn from the posterior using the COBE/FIRAS and PIXIE-type likelihoods, respectively.
    \label{fig:CosmoALP_bayesian}}
\end{figure}

\Cref{fig:CosmoALP_bayesian} shows the Bayesian equivalent of \cref{fig:results}. The key difference here compared to \cref{fig:results} is that both the priors and the posterior are plotted in \cref{fig:CosmoALP_bayesian}. Note that whilst these panels do show individual points, they do not show all points sampled. The points shown are an equal number of equal-weight draws from each distribution. This is in contrast to \cref{fig:results}, which is based on profiling the likelihoods of all sample points within fixed-width bins, and interpolating the resulting maximum likelihood between them.

For a likelihood $\mathcal{L}$ and prior $\pi$, one may compute a Bayesian evidence (or marginal likelihood) $\mathcal{Z}$, posterior distribution $\mathcal{P}$ from Bayes theorem, and Kullback--Leibler divergence $\mathcal{D}_\text{KL}$ between the prior and posterior distribution~\cite{Trotta:2008qt}:
\begin{equation}
    \mathcal{P} = \frac{\mathcal{L}\pi}{\mathcal{Z}} \qquad \mathcal{Z} = \int \mathcal{L} \pi d\theta = \left\langle \mathcal{L} \right\rangle_\pi \qquad \mathcal{D}_\text{KL} = \int \mathcal{P} \ln\frac{\mathcal{P}}{\pi} d\theta  = \left\langle \ln \frac{\mathcal{P}}{\pi}\right\rangle_\mathcal{P}
\end{equation}
In order to quantitatively compare the two models, \LCDM and \LCDM+ALPs, we compute the Bayesian evidence, $\mathcal{Z}$ for the different prior definitions:
\begin{equation}
    \begin{aligned}
    \ln \mathcal{Z}_\text{ALPs}^\text{box} &= -1039.09\pm 0.14 \\
    \ln \mathcal{Z}_\text{ALPs}^\text{Primakoff} &= -1038.12\pm 0.14 \\
    \ln \mathcal{Z}_{\Lambda\text{CDM}} &= -1037.11\pm 0.14
    \end{aligned}
\end{equation}
\LCDM is thus marginally preferred to the ALP model in a Bayesian sense by $\Delta\ln\mathcal{Z}^{\text{box}} = -1.98\pm 0.20$ or $\Delta\ln\mathcal{Z}^\text{Primakoff}=-1.02\pm 0.20$ log units. The error bars in this section all arise from the nested sampling procedure~\cite{Skilling:2006,nested_sampling} as a result of the probabilistic integration used for estimating the evidence.

We can see where the difference in Bayesian evidence arises by using the Occam's razor equation to decompose the evidence into a goodness of fit (average log-likelihood) and Occam penalty (Kullback--Leibler divergence)~\cite{Hergt:2021qlh}.
By taking logarithms and rearranging Bayes theorem before posterior averaging, we recover the Occams razor equation, which in our case decomposes the evidences as
\begin{equation}
    \begin{aligned}
        \ln \mathcal{Z} &= \phantom{-}\left\langle\ln\mathcal{L}\right\rangle_\mathcal{P} &-\,& \mathcal{D}_\text{KL} \\
        \ln \mathcal{Z}_\text{ALPs}^\text{box} &= -1012.27 &-\,& 26.82 \\
        \ln \mathcal{Z}_\text{ALPs}^\text{Primakoff} &= -1012.27 &-\,& 25.85 \\
        \ln \mathcal{Z}_{\Lambda\text{CDM}} &= -1012.38 &-\,& 24.72
    \end{aligned}
\end{equation}
These results demonstrate that all models in a Bayesian sense fit the data equally well, with $\Delta\left\langle\ln\mathcal{L}\right\rangle_\mathcal{P}=0.11\pm0.07$ relative to \LCDM, consistent with zero. The ALPs are penalised by having additional constrained parameters, with an additional compression of $\Delta\mathcal{D}_\text{KL}^\text{box} = 2.09\pm 0.19$ or $\Delta \mathcal{D}_\text{KL}^\text{Primakoff}=1.12\pm 0.19$, i.e.\ the parameter space contained in the posterior is a factor of $\exp(\Delta\mathcal{D}_\text{KL}^\text{box})= 8.3\pm1.5$ or $\exp(\Delta\mathcal{D}_\text{KL}^\text{Primakoff})=3.1\pm0.6$ smaller than the prior. This compression factor measures how well the data constrain the model. Note that employing the more stringent Primakoff prior rules out a little more than half the parameter space in advance, and therefore reduces the Occam penalty.
If the COBE/FIRAS data set is replaced with the PIXIE forecast, the only change is an increase in parameter space compression by approximately a factor of $\e$: $\Delta\mathcal{D}_\text{KL}^\text{box}=3.08\pm0.19$ and $\Delta\mathcal{D}_\text{KL}^\text{Primakoff}=2.11\pm0.19$, i.e.\ the posterior is a factor of $\exp(\Delta\mathcal{D}_\text{KL}^\text{box})= 22.2\pm4.2$ or $\exp(\Delta\mathcal{D}_\text{KL}^\text{Primakoff})=8.4\pm1.6$ smaller than the prior. The Likelihood contribution to the evidence remains the same.

\begin{figure}
    \centering
    \includegraphics[width=6in]{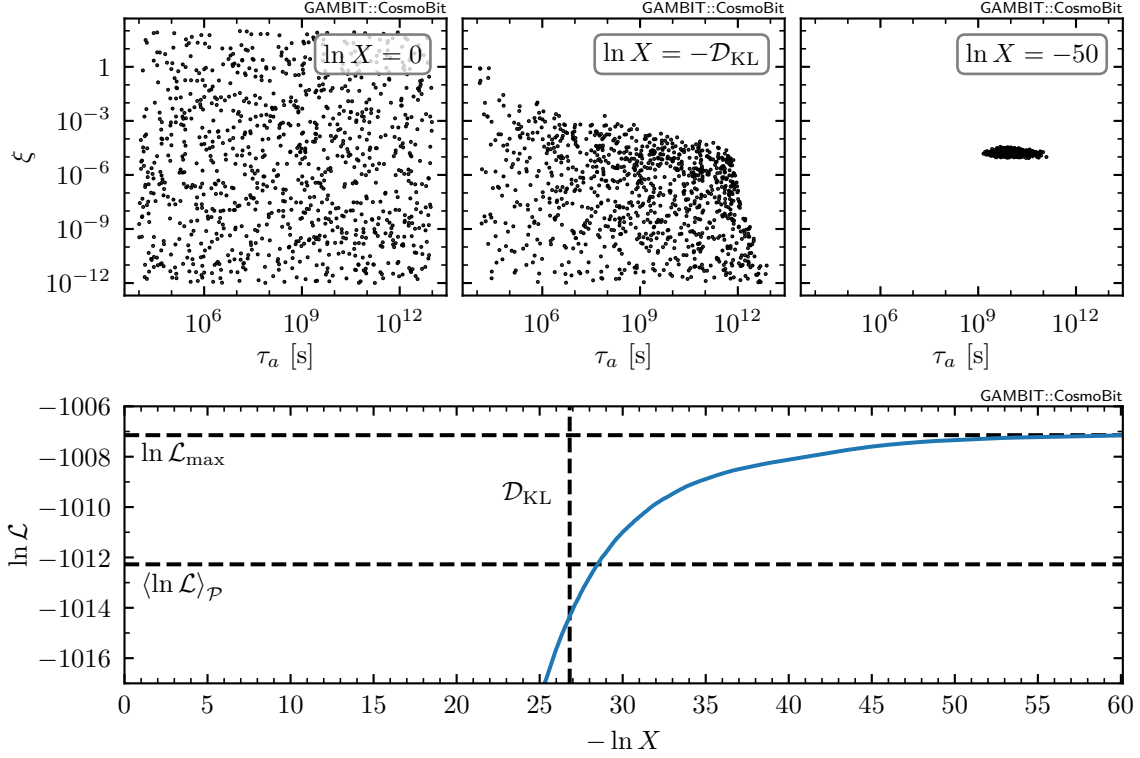}
    \caption{
        \label{fig:CosmoALP_bayesian_phase}
        \textit{Top}: Live point locations in the $(\tau_a,\xi)$ plane during different stages of nested sampling~(NS). The $i$th NS iteration may be parametrised by the log-prior volume ${\ln X \approx -i/n_\text{live}}$. Three critical stages are shown: the initial (non-Primakoff) prior points when $\ln X = 0$, at the ``typical set'' where $\ln X =-\mathcal{D}_\text{KL}$, and at a much lower value around the point where NS terminates according to our selected tolerance (which in this case is around the best-fit point). The prior points have large volume and low likelihood, whereas the peak has high likelihood and low volume. The Bayesian ``sweet spot'' (i.e.\ majority contribution to the evidence) occurs around the typical set with approximate parameter volume  given by $\e^{\mathcal{D}_\text{KL}}$. The higher-likelihood phase therefore occupies a parameter region $\approx 10^7$ times smaller than the usual posterior, so is not preferred in a Bayesian sense and does not appear in \cref{fig:CosmoALP_bayesian}.
        \textit{Bottom}: Phase plot of log-likelihood against log-volume, showing the average log-likelihood $\langle\ln\mathcal{L}\rangle_\mathcal{P}$ over the posterior $\mathcal{P}$, the Kullback--Leibler divergence $\mathcal{D}_\text{KL}$, and the maximum log-likelihood value $\ln\mathcal{L}_\text{max}$. The high likelihood peak can be identified as beyond the region where there is a small inflection point at $\ln X\approx -42$.
        All of this information is extractable in post-processing with the NS Python package \textsf{anesthetic}~\cite{anesthetic}.
    }
\end{figure}

\afterpage{\clearpage}

\Cref{fig:CosmoALP_bayesian_phase} shows that using the full scope of the nested sampling run, we can gain insight into the high likelihood region shown in \cref{fig:results}. This plot showcases the Bayesian interplay between likelihood and parameter volume.

\section{Conclusions}

Axion-like particles (ALPs) are well-motivated candidates for light new physics and arise generically in many extensions of the Standard Model. In the present work we studied ALPs with masses in the keV--MeV range. We focused on ALPs with couplings to photons small enough to be inaccessible to accelerator experiments, but large enough to ensure that the ALPs decay before recombination and therefore do not contribute to the observed dark matter abundance. In this coupling range, ALPs can affect various cosmological and astrophysical observables, which makes it possible to search for them and to constrain their properties.

A central aspect of our analysis is that we did not assume that ALPs enter into thermal equilibrium in the early universe, such that their abundance is essentially an independent parameter. Nevertheless, one can place a lower bound on the abundance from the irreducible freeze-in production, and we provided a detailed calculation of this abundance as a function of the ALP mass and reheating temperature. At the same time, BBN and CMB constraints (both from anisotropies and spectral distortions) place tight upper bounds on the ALP abundance, typically requiring that these particles constitute only a tiny fraction of the dark matter abundance even before they decay.

To perform a global analysis of the ALP parameter space, we interfaced the GAMBIT framework with a number of external codes in order to combine state-of-the-art likelihood calculations for all relevant data sets. Most notably, for the first time we combine detailed calculations of the effects of photodisintegration of light elements with constraints on CMB spectral distortions, both of which give highly relevant constraints on ALP models. Our setup makes it possible to scan simultaneously over the ALP model parameters, the \LCDM parameters, and various nuisance parameters, in order to identify the most interesting regions of parameter space (see \cref{fig:results}).

In particular, the decays of ALPs into photons can modify both the abundances of light elements and the baryon-to-photon ratio $\eta_b$. As a result, we find that $[\ce{D}/\ce{H}]$ and $\Omega_b h^2$ are less tightly correlated in our model than in \LCDM (see \cref{fig:observables}). This makes it possible to achieve a slightly better fit to observations in our model than in \LCDM, although the Bayesian evidence does not prefer the ALP model over standard \LCDM. However, we did not find any allowed parameter regions in our model that would address the $\ce{^7Li}$ problem.

Compared to previous results, we note that leaving the \updated{nonthermal abundance as a free parameter} opens up parameter space for higher ALP masses, which are otherwise excluded by cosmological constraints (see \cref{fig:summary_plot}). Nonetheless, bounds on the decays of ALPs produced in \sn still provide relevant constraints in that region. This emphasises the importance of an analysis framework that allows for the inclusion of all relevant parameters and complementary constraints from astrophysics -- both of which can be achieved with global-fitting tools such as \gambit.

Our findings also highlight the fact that measurements of the spectrum of the CMB from over 20 years ago can provide some of the leading cosmological constraints for ALPs with masses around the MeV scale. In particular, these constraints are stronger than those from the photodisintegration of light elements after BBN, which only constrain ALP masses $\gtrsim \SI{5}{\MeV}$.
More accurate measurements from a future satellite mission such as PIXIE therefore promise substantially stronger constraints and the potential to discover weakly coupled MeV-scale ALPs via spectral distortions.

To conclude, we emphasise that our analysis was focused on a fairly restrictive set of ALP models, in which the ALP couplings to electrons and hadrons are assumed to be negligible.
While this is a common assumption in the literature, there has been substantial, recent progress in understanding the effects of ALP-matter couplings on cosmology~\cite{1604.08614,1808.07430,2007.01873,2109.12088}.
This research programme goes hand in hand with a detailed exploration of the importance of ALP-matter coupling for supernovae such as SN1987A~\cite{2008.04918,2008.11741,2109.03244}.
Moreover, ALPs with hadronic couplings have been proposed as a possible solution of the $\ce{^7Li}$ problem~\cite{Goudelis:2015wpa}.
It will therefore be very interesting to extend our setup and explore the interplay between such ALP models and observations, and the resulting constraints on more general ALP parameter spaces.

\acknowledgments

{\small
We thank Deanna Hooper, Lloyd Knox, Julien Lesgourgues, Kai Schmidt-Hoberg, Edoardo Vitagliano, and all members of the GAMBIT community for discussions. We are thankful to Sebastian Wild for collaboration in the early stages of this work.
We are furthermore very grateful to Marco Hufnagel for help with ACROPOLIS, to Matteo Lucca and Nils Sch{\"o}neberg for help with the spectral distortions code, and to Joerg Jaeckel and Pedro~C.\ Malta for helpful discussions on Ref.~\cite{1702.02964} and letting us inspect the associated software code.
CB received financial support from the Australian Research Council Discovery Projects DP180102209 and DP210101636.
TEG, FK and PSt were funded by the Deutsche Forschungsgemeinschaft (DFG) through the Emmy Noether Grant No.\ KA 4662/1-1.
SH was supported by the Alexander von Humboldt Foundation and the German Federal Ministry of Education and Research.
DJEM is supported by an Ernest Rutherford Fellowship from the Science and Technologies Facilities Council (UK).
PSc acknowledges funding support from the Australian Research Council under Future Fellowship FT190100814.
PSt and SH thank FK for hospitality at RWTH~Aachen, where part of this work was done.
We thank PRACE for access to the Joliot-Curie HPC cluster at CEA.
This work was performed using the Cambridge Service for Data Driven Discovery (CSD3), part of which is operated by the University of Cambridge Research Computing on behalf of the STFC DiRAC HPC Facility (\url{www.dirac.ac.uk}). The DiRAC component of CSD3 was funded by BEIS capital funding via STFC capital grants ST/P002307/1 and ST/R002452/1 and STFC operations grant ST/R00689X/1. DiRAC is part of the National e-Infrastructure.
This work used the Scientific Compute Cluster at GWDG, the joint data centre of Max Planck Society for the Advancement of Science (MPG) and University of {G\"ottingen}.
This article made use of \textsf{pippi} v2.1 \cite{pippi}.
}

\appendix

\section{GAMBIT implementation}

The computations and simulations perfomed in this work were done using the global fitting framework \gambit. In this appendix we document the modifications and additions we implemented for the purpose of this study.
For full details on the structure of \gambit, we refer the reader to the \gambit core paper~\cite{gambit}.
Further details on the \cosmobit extension can be found in Refs.~\cite{CosmoBit,CosmoBit_numass}.

\subsection{New models} \label{app:models}

The new models added to \gambit for this study are a generic parametrisation of a cosmological ALP, \doublecrosssf{GeneralCosmoALP}{GeneralCosmoALP}, two simplified ALP models \doublecrosssf{CosmoALP}{CosmoALP} and \doublecrosssf{CosmoALP\_gg\_tau}{CosmoALP_gg_tau}, a re-parametrisation of \LCDM, \doublecrosssf{LCDM\_zreio}{LCDM_zreio}, two new models for non-standard radiation, \doublecrosssf{etaBBN\_rBBN\_dNurBBN}{etaBBN_rBBN_dNurBBN} and \doublecrosssf{rBBN\_dNurBBN}{rBBN_dNurBBN}, and two models of nuisance parameters, \doublecrosssf{cosmo\_nuisance\_BK15}{cosmo_nuisance_BK15} and \doublecrosssf{cosmo\_nuisance\_SpectralDistortions}{cosmo_nuisance_SpectralDistortions}.

Various models of ALPs were already implemented in \gambit in Ref.~\cite{Axions}, including the \textsf{GeneralALP} model, which sits at the top of the model hierarchy.
We generalised this base model even more to take into account cosmological effects, such as the thermal abundance of ALPs and the reheating temperature of the Universe, introducing a new ``parent model'' of \textsf{GeneralALP} named \doublecrosssf{GeneralCosmoALP}{GeneralCosmoALP}. We also added two other ``child models'' of \doublecrosssf{GeneralCosmoALP}{GeneralCosmoALP}: \doublecrosssf{CosmoALP}{CosmoALP} and \doublecrosssf{CosmoALP\_gg\_tau}{CosmoALP_gg_tau}.
As ALPs may decay anytime before neutrino decoupling and the present day, translation functions are provided to the decaying DM model \textsf{DecayingDM\_photon}, assuming that the primary decay mode is via photons.
A ``friendship relation'' between these models ensures that energy injection to the CMB is taken into account in models with short ALP lifetimes.

\begin{description}
 \item[\textbf{\textsf{GeneralCosmoALP}}: \label{GeneralCosmoALP}] \term{gagg, gaee, gaN, fa, ma0, Tchi, beta, thetai, f0\_thermal, T\_R}

 Generalization of the \textsf{GeneralALP} model from Ref.~\cite{Axions}, with the addition of the thermal ALP abundance $\xi$~(the \texttt{f0\_thermal} parameter) and reheating temperature \TR.

 \item[\textbf{\textsf{CosmoALP}}: \label{CosmoALP}] \term{Cagg, fa, ma0, thetai, f0\_thermal, T\_R}

 Child model of \doublecrosssf{GeneralCosmoALP}{GeneralCosmoALP}, where the ALP-electron coupling and ALP-nucleon coupling are neglected, $g_{ae}=g_{aN}=0$, and with temperature-independent mass, i.e. $\beta=0$ and $T_\chi$ a large and irrelevant value. The ALP-photon coupling $g_{a\gamma}$ is replaced by its dimensionless equivalent $C_{a\gamma}$.

 \item[\textbf{\textsf{CosmoALP\_gg\_tau}}: \label{CosmoALP_gg_tau}] \term{tau, fa, ma0, Tchi, beta, thetai, f0\_thermal, T\_R}

 Child model of \doublecrosssf{GeneralCosmoALP}{GeneralCosmoALP} where the ALP-electron coupling and ALP-nucleon coupling are neglected, $g_{ae}=g_{aN}=0$, and the ALP-photon coupling is replaced by the ALP lifetime $\tau_a$.

\end{description}

\noindent The standard parametrisation of \LCDM in \gambit includes the baryon density $\omega_b$, the DM density $\omega_{\text{DM}}$, the Hubble constant $H_0$, the CMB temperature \TCMB, and the optical depth at reionisation $\tau_\text{reio}$. However, some of the external libraries we use, e.g.\ \textsf{CLASS}, internally use the redshift at reionisation $z_\text{reio}$ instead of $\tau_\text{reio}$. Typically obtaining $z_\text{reio}$ from $\tau_\text{reio}$ is not an issue, but in the presence of energy injection at the time of reionisation this is not trivial. For this reason we introduce a new model, \doublecrosssf{LCDM\_zreio}{LCDM_zreio}.

\begin{description}

\item[\textbf{\textsf{LCDM\_zreio}}: \label{LCDM_zreio}] \term{omega\_b, omega\_cdm, H0, T\_cmb, z\_reio}

Re-parametrisation of the \LCDM model with the redshift at reionisation $z_\text{reio}$ instead of the optical depth at reionisation $\tau_\text{reio}$.

\end{description}

\noindent In \gambit, all non-standard radiation content that modifies the Hubble rate is parametrised by the five parameter model \textsf{etaBBN\_rBBN\_rCMB\_dNurBBN\_dNurCMB} and children thereof, cf.\ Ref.~\cite[appendix~B.4]{CosmoBit}.
For this analysis we have implemented two additional models, \doublecrosssf{etaBBN\_rBBN\_dNurBBN}{etaBBN_rBBN_dNurBBN} and \doublecrosssf{rBBN\_dNurBBN}{rBBN_dNurBBN}, which allow for scenarios in which the modifications to \Neff at BBN are free model parameters that fully determine the respective values at recombination.

\begin{description}

\item[\textbf{\textsf{etaBBN\_rBBN\_dNurBBN}}:\label{etaBBN_rBBN_dNurBBN}] \term{dNur\_BBN, r\_BBN, eta\_BBN}.

Parametrisation of non-standard radiation contect with $\eta_\text{BBN}$, the baryon-to-photon ratio at the end of BBN; $\Delta N_\text{ur, BBN}$, the ultra-relavistic degrees of freedom at the end of BBN; and $r_{\nu,\text{BBN}}$, the neutrino temperature ratio at the end of BBN. A translation function is provided to map these parameters to those of \textsf{etaBBN\_rBBN\_rCMB\_dNurBBN\_dNurCMB}, requiring the new capability \cpp{Neff\_evolution}, defined below.

\item[\textbf{\textsf{rBBN\_dNurBBN}}:\label{rBBN_dNurBBN}] \term{dNur\_BBN, r\_BBN}

Child model of \doublecrosssf{etaBBN\_rBBN\_dNurBBN}{etaBBN_rBBN_dNurBBN}, where $\eta_\text{BBN}$ is inferred from the baryon-to-photon ratio today $\eta_0$, and thus the translation depends on the new capability \cpp{eta\_ratio}, defined below.

\end{description}

\noindent Lastly, we introduced nuisance parameter models to include the systematic uncertainties of various experiments whose likelihoods are made available by \textsf{MontePython}. The nuisance model for the BICEP/Keck~Array was updated to include the likelihood for the BK15 results~\cite{1810.05216} in the model \doublecrosssf{cosmo\_nuisance\_BK15}{cosmo_nuisance_BK15}. Additionaly, as this work is the first introduction of CMB spectral distortions in \gambit (cf.\ \cref{sec:sd}), we implemented a model to include the nuisance parameters for the COBE/FIRAS and PIXIE experiments: \doublecrosssf{cosmo\_nuisance\_SpectralDistortions}{cosmo_nuisance_SpectralDistortions}.

\begin{description}

\item[\textbf{\textsf{cosmo\_nuisance\_BK15}}: \label{cosmo_nuisance_BK15}]\term{BBbetadust, BBbetasync BBdust, BBsync, BBalphadust, BBTdust, BBalphasync, BBdustsynccorr, EEtoBB\_dust, EEtoBB\_sync, Delta\_dust, Delta\_sync, gamma\_corr, gamma\_95, gamma\_150, gamma\_220}

Extension of the existing \textsf{cosmo\_nuisance\_BK14} model including additional systematics.

\item[\textbf{\textsf{cosmo\_nuisance\_SpectralDistortions}}: \label{cosmo_nuisance_SpectralDistortions}] \term{sd\_delta\_T,         sd\_T\_D, sd\_beta\_D, sd\_A\_D, sd\_T\_CIB, sd_beta\_CIB, sd\_A\_CIB, sd\_alpha\_sync, sd\_omega\_sync, sd\_A\_sync, sd\_T\_e, sd\_EM, sd\_nu\_p\_spin, sd\_A\_spin, sd\_A\_CO, sd\_y\_reio\_nuisance}

The spectral distortion nuisance model contains parameters that take into account various uncertainties on the COBE/FIRAS and PIXIE experiments.

\end{description}

\subsection{New capabilities provided in \cosmobit and \darkbit} \label{app:capabilities}

In addition to the new models introduced for this study, we also added new capabilities, and modified existing ones, to work with the ALP models. Most of the new module functions that provide these capabilities were added to \cosmobit, as they had to do with the cosmological implications of ALPs, but one was added to \darkbit, as it deals with late-time astrophysical constraints. The full list of new capabilities, along with the module functions that provide them and their dependencies and backend requirements, can be seen in \cref{tab:capabilities_axions}. We give a quick summary of the most important ones below.

\begin{table*}[tp]
\centering
  \caption{Capabilities added to \cosmobit and \darkbit associated with this study. Here, \cpp{map_str_dbl} is a \cpp{typedef} for \cpp{std::map<std::string,double>}, \cpp{map_str_int} for \cpp{std::map<std::string,int>} and \cpp{[pd]} is shorthand for \cpp{photodisintegration}. All functions are contained in \cosmobit except for \cpp{calc_PhotonFluence_SN1987A_Decay}, which is part of \darkbit.}
\scriptsize
   \makebox[\linewidth]{
   \begin{tabular}{p{4.4cm} p{6.8cm} p{5cm} }

    \toprule
   \textbf{Capability}  & \textbf{Function} (\textbf{type}) & \textbf{Dependencies [type] / \newline Backend reqs [type (args)]} \\ \midrule

    \cpp{lifetime}
      &\cpp{lifetime\_ALP\_agg} (\cpp{double}) & {\cpp { } }  \\
   \midrule

    \cpp{minimum\_abundance}
      &\cpp{minimum\_abundance\_ALP\_analytical} (\cpp{double}) \newline \cpp{minimum\_abundance\_ALP\_numerical} (\cpp{double})& {\cpp { } }  \\
   \midrule

    \cpp{minimum\_fraction}
      &\cpp{minimum\_fraction\_ALP} (\cpp{double}) & {\cpp {minimum\_abundance [double]} }  \\
   \midrule

    \cpp{DM\_fraction}
    &\cpp{DM\_fraction\_ALP} (\cpp{double}) & {\cpp {minimum\_fraction [double]} }
    \newline {\cpp {lifetime [double]} }
    \newline {\cpp {RD\_oh2 [double]} } \\
   \midrule

    \cpp{total\_DM\_abundance}
      &\cpp{total\_DM\_abundance\_ALP} (\cpp{double}) & {\cpp {DM\_fraction [double]} } \\
   \midrule

    \cpp{external\_dNeff\_etaBBN}      &\cpp{compute\_dNeff\_etaBBN\_ALP} (\cpp{map\_str\_dbl}) & {\cpp {total\_DM\_abundance [double]} }
    \newline {\cpp {lifetime [double]} } \\
   \midrule

    \cpp{eta_ratio} & \cpp{eta_ratio_ALP} (\cpp{double}) & \cpp{external_dNeff_etaBBN [double]} \\
   \midrule

    \cpp{Neff_evolution} & \cpp{Neff_evolution_ALP} (\cpp{map_str_dbl}) & \cpp{external_dNeff_etaBBN [double]}  \\
   \midrule

    \cpp{tau_reio} & \cpp{get_tau_reio_classy} (\cpp{double}) & \cpp{class_get_tau_reio [double ()]}  \\
   \midrule

    \cpp{z_reio} & \cpp{get_z_reio_classy} (\cpp{double}) & \cpp{class_get_z_reio [double ()]} \\
   \midrule

    \cpp{H_at_z} & \cpp{get_H_at_z_classy} (\cpp{daFunk::Funk}) & \cpp{class_get_Hz [double (double)]}  \\
   \midrule

   \cpp{time_at_z} & \cpp{get_time_at_z_classy} (\cpp{daFunk::Funk}) & \cpp{class_get_tz [double (double)]}  \\
   \midrule

   \cpp{age_universe} & \cpp{get_age_universe_from_time_at_z} (\cpp{double}) & \cpp{time_at_z [daFunk::Funk]} \\
   \midrule

   \cpp{primordial_abundances_BBN} & \cpp{compute_primordial_abundances_BBN} (\cpp{BBN_container}) & \cpp{AterBBN_Input [map_str_dbl]}\newline \cpp{call_nucl_err [int (map_str_dbl, double, double)]} \newline \cpp{call_nucl [int (map_str_dbl,double)]} \newline \cpp{get_NNUC [size_t ()]} \newline \cpp{get_abund_map_AlterBBN [map_str_int ()]} \\
   \midrule

   \cpp{primordial_abundances} & \cpp{primordial_abundances} (\cpp{BBN_container}) \newline\cpp{primordial_abundances_decayingDM} (\cpp{BBN_container}) & \cpp{primordial_abundances_BBN [BBN_container]}\newline \cpp{set_input_params [void (bool,int,int,double)} \newline \cpp{abundance_[pd]_decay [void  (double*,...)]} \\
   \midrule

   \cpp{PhotonFluence_SN1987A_Decay}  & \cpp{calc_PhotonFluence_SN1987A_Decay} (\cpp{double}) &  \\
   \midrule

   \end{tabular}
  }
  \label{tab:capabilities_axions}
\end{table*}

A few of the newly added capabilities involve computing properties of the ALPs that can be derived from the model parameters. The capability \cpp{lifetime} computes the ALP lifetime as a function of the mass and coupling. The \cpp{minimum\_abundance} capability computes the minimal freeze-in abundance of ALPs, produced via Primakoff processes, either analytically or numerically (by reading in tables pre-computed with \micro), and \cpp{minimum\_fraction} computes the fraction of the total abundance of dark matter constituted by the minimal freeze-in abundance. The \cpp{total\_DM\_abundance} is the total abundance of ALPs, produced by realignment and freeze-in, and \cpp{DM\_fraction} provides the fraction of the total DM abundance constituted by ALPs (assuming no ALP decays).

In decaying DM models such as ours, the baryon-to-photon ratio at BBN $\eta_{\rm{BBN}}$, the neutrino-to-photon temperature ratio $r_\nu$, and the number of ultra-relativistic degrees of freedom at recombination $\Delta N_{\rm{eff}}$, are provided by the capability \cpp{external_dNeff_etaBBN}. These values are returned as a \cpp{std::map}, and are can be accessed individually using the capabilities \cpp{eta_ratio} for $\eta_{\rm{BBN}}$ and \cpp{Neff_evolution} for both $r_\nu$ and $\Delta N_{\rm{eff}}$. Details of these computations can be found in Appendix~\ref{appendix:neff_impl}.

As mentioned above in \ref{app:models}, we have defined a re-parametrisation of the \LCDM model in \gambit to more closely match the parametrisation in \textsf{CLASS}. Along with the new model, a few capabilities are defined to provide useful quantities. \cpp{tau_reio} and \cpp{z_reio} extract the values of the optical depth and redshift of reionisation from \textsf{CLASS}. The capabilities \cpp{H_at_z} and \cpp{time_at_z} return the Hubble parameter and time since the Big Bang from \textsf{CLASS} as \cpp{daFunk} functions of redshift (see \cite{darkbit} for more information on the \cpp{daFunk} library). Lastly \cpp{age_universe} computes the age of the Universe at a given redshift with \textsf{CLASS}.

To take into account the effects of the photodisintegration of light elements due to the EM spectrum injected by the ALP decays, we have restructured the capabilities relating to BBN. The capability \cpp{primordial_abundances_BBN} replaces the old \cpp{BBN_abundances} capability, which computes the abundances after BBN using \textsf{AlterBBN}. The new capability \cpp{primordial_abundances} computes the abundances at the present time. Two module functions provide this capability, \cpp{primordial_abundances}, which simply returns the abundances unchanged from BBN, and \cpp{primordial_abundances_decayingDM}, which uses \textsf{ACROPOLIS} to modify the computed abundances at BBN in order to include the effects of photodisintegration.

Finally, the only new capability added to \darkbit is \cpp{PhotonFluence_SN1987A_Decay}, which computes the $\gamma$ fluence from the supernova SN1987A as a result of the decay of ALPs into photons.

\subsection{New and updated backend interfaces}
\label{app:backends}

\gambit provides interfaces to a large collection of external tools, known as backends. The backend interfaces are documented along with the modules that use them~\cite{ColliderBit, CosmoBit, darkbit, FlavBit, RHN, SDPBit}. Backend interfaces can also be auto-generated~\cite{GUM}. For this analysis we have updated some of the backends reported in \cite{CosmoBit} to their latest version, such as \textsf{CLASS 3.1}~\cite{Blas:2011rf} and \textsf{MontePython 3.5}~\cite{brinckmann2018montepython}. The updated version of \textsf{CLASS} brings along the possibility of including spectral distortions, as described in \cref{sec:sd}, which also benefits from the new version of \textsf{MontePython}, as it comes with more likelihoods, including COBE/FIRAS and a mock PIXIE-like likelihood. Furthermore, we have modified the interface to \textsf{AlterBBN}~\cite{Arbey:2011nf,Arbey:2018zfh} in order to allow the computation of abundances without uncertainties, as well as to return three values of the abundances, corresponding to the central, high and low values computed by \textsf{AlterBBN}.

The only absolutely new backend interface implemented for this study is to the external tool \textsf{ACROPOLIS 1.2.1} \cite{Depta:2020mhj}. Written in \textsf{Python}, \textsf{ACROPOLIS} computes the effects of photodisintegration of light elements on pre-computed abundances. In addition to the light element abundances, \textsf{ACROPOLIS} takes as input the mass of the decaying particle, $m_a$, its lifetime $\tau_a$, the ratio of the number density of the decaying particle today to that of photons, $N_{0,a}$, and the branching fractions to electrons and photons, respectively. From this set of input parameters, the backend convenience function \cpp{abundance_photodisintegration_decay} performs various transformations on the input and output abundances and covariance matrix, and returns the final abundances after photodisintegration, and their associated covariance matrix.

The light element abundances in \textsf{ACROPOLIS} are taken as absolute abundances, as opposed to the treatment in \gambit and \textsf{AlterBBN}, where they are treated as ratios over the hydrogen abundance, $r_i = Y_i/Y_H$. Hence before feeding the abundances to \textsf{ACROPOLIS}, they are modified accordingly. Furthermore, the abundances computed by \textsf{AlterBBN} include the late-time decays of $\ce{^3H}$ and $\ce{^7Be}$. As these happen after photodisintegration, and are in fact automatically added by \textsf{ACROPOLIS}, we revert the $\ce{^3H}$ and $\ce{^7Be}$ decays when converting to absolute abundances for sending to \textsf{ACROPOLIS}. Conversely, before returning the final abundances, they are converted back to ratios over $Y_H$.

\textsf{ACROPOLIS} takes the abundances as a fixed-size array ordered as $\{n, p, D, \ce{^3H}, \ce{^3He}, \ce{^4He}, \ce{^6Li}, \ce{^7Li}, \ce{^7Be}\}$ and, along with the input parameters, computes a transfer matrix $\mathbf{M}$, which provides the abundances after photodisintegration as $\mathbf{Y}_f = \mathbf{M} \mathbf{Y}_i$. Besides the computation of the final abundances, this transfer matrix is used to propagate the covariance matrix as
\begin{equation}
 \mathbf{\Sigma}_f = (\mathbf{J}_2  \mathbf{M}  \mathbf{J}_1)  \mathbf{\Sigma}_i  (\mathbf{J}_2  \mathbf{M}  \mathbf{J}_1)^T,
\end{equation}
where $\mathbf{J}_1$ is the Jacobian of the transformation from ratios over $Y_H$ to absolute abundances, and $\mathbf{J}_2$ the converse. These Jacobians are different because the former includes the subtraction of late-time decays, whereas these are not included in the latter, but rather in the transfer matrix $\mathbf{M}$.

Lastly, there are few internal options in \textsf{ACROPOLIS} that determine the precision and speed of the computations, among others. These can be provided as options in the \YAML file, which can be seen below, including three configurations for default, fast and aggressive settings. The backend convenience function \cpp{set_input_params} ensures that the given options are passed to the relevant \textsf{ACROPOLIS} modules.

\begin{lstyaml}
  # Options for ACROPOLIS
  - capability: primordial_abundances
    function: primordial_abundances_decayingDM
    options:
      verbose: false
      NE_pd: 30 # default: 150, fast: 75, aggresive: 30
      NT_pd: 10 # default: 50, fast: 25, aggresive: 10
      eps: 1e-1 # default: 1e-3, fast: 1e-2, aggresive: 1e-1
\end{lstyaml}

\section{Details of implementation of entropy injection between BBN and recombination}\label{appendix:neff_impl}

The evolution of the photon temperature including photon production through ALP decays $ \ax \to \gamma\gamma $ is given by \cref{eq:dT_dt_gamma}. It is important to note that the functions $ n_a(t) $ and $ H(t) $ in \cref{eq:dT_dt_gamma} both have an implicit dependence on the photon temperature $ T_\gamma $ and neutrino temperature $ T_\nu $. Both temperatures affect the energy density in radiation $ \rho_\text{rad} $, see \cref{eq:rad_dens_general}, which in turn affects the expansion rate $ H(t) $ and the scale factor $ \scalef(t) $ entering $ n_a(t) $. Self-consistently integrating \cref{eq:dT_dt_gamma} therefore requires taking the implicit temperature dependence into account. Doing so can, however, lead to numerical instabilities. In order to circumvent these complications, we take an alternative approach: instead of solving \cref{eq:dT_dt_gamma} only once, while respecting the implicit dependence of $ H(t) $ and $ n_a(t) $ on $ T_\gamma $, we solve this equation several times while neglecting the changes in $ H(t) $ and $ \scalef(t) $ due to the changes in $ T_\gamma $. After each iteration $ H(t) $ and $ \scalef(t) $ are updated and the differential equation is solved again. This means that in the $i$th iteration, the differential equation is given by
\begin{equation}
\frac{\dd T^{(i)}_\gamma}{\dd t} = \frac{15}{4 \pi^2} \frac{m_a n_a^{(i-1)}(t)}{\tau_a} \frac{1}{(T^{(i)}_\gamma)^3} - H^{(i-1)}(t) T^{(i)}_\gamma \,,
\label{eq:iterative_DE}
\end{equation}
where $ H^{(i-1)}(t) $ and $ n_a^{(i-1)}(t) $ are derived assuming the photon temperature of the previous iteration $ T^{(i-1)}_\gamma $. This procedure is repeated until $ T_\gamma $ has converged, i.e.\ $ T^{(i)}_\gamma(t) = T^{(i-1)}_\gamma(t) $.

\subsection{Numerical implementation}
Here we describe the numerical implementation of this procedure. We create a look-up table of the photon temperature $ T_\gamma(t) $, the neutrino temperature $ T_\nu(t) $, the Hubble rate $ H(t) $, as well as the natural logarithm of the scale factor $ \ln \scalef(t) $, which enter $ n_a(t) $ as given by \cref{eq:na_t}.\footnote{In the scenario studied in this paper, the neutrino temperature can be derived through \cref{eq:Tnu_t} such that there would be no need to include $ T_\nu(t) $ in this table, as it can be directly derived through $ \ln \scalef(t) $. We include this, however, to allow for easy generalisation to scenarios in which the neutrino temperature is also modified.}

When this table is initialised ($ i=0 $), it has logarithmic spacing in time between $ t_\text{ini} $ and $ t_\text{fin} $. Furthermore, we assume that there are no modifications to the photon temperature and neutrino temperature, meaning that initially the effective number of relativistic degrees of freedom $ g_\ast(T) $ is assumed to be constant in the temperature range of interest. In this case $ H = \dot{\scalef}/\scalef \sim T_\gamma^2 $ and $ T_\gamma \sim 1/\scalef $ such that the initial Hubble rate $ H^{(0)}(t) $ is given by
\begin{equation}
  H^{(0)}(t) = \frac{1}{2\,t} \, .
\end{equation}
Using the right-hand side of \cref{eq:Hubble_rad}, the initial photon temperature $ T_\gamma^{(0)}(t) $ is calculated from $ H^{(0)}(t) $ and the initial neutrino temperature is given by assuming instant decoupling $ T^{(0)}_\nu = \left(4/11\right)^{1/3} T^{(0)}_\gamma $. The last column of the table $ \ln a^{(0)}(t) $ is initialised via
\begin{equation}
  \ln \scalef^{(0)}(t) = \int_{t_\text{ini}}^{t} \dd t' \, H^{(0)}(t') \, .
\end{equation}

After \cref{eq:iterative_DE} is solved the table has to be updated in order to be self-consistent. In particular, as $ T_\gamma(t) $ is modified, $ H(t) $ is modified as well. Moreover, it is important to note that the constant column in this table is not the time $ t $ but the scale factor or its natural logarithm $ \ln \scalef(t) $. Hence, once the column containing $ H $ is updated by using the respective values of $ T_\gamma $ (and $ T_\nu $) after solving the differential equation, the column containing the time coordinate $ t $ has to be updated as well by using $ H = \dd \ln \scalef/\dd t $, such that $ t $ becomes a function of $ \ln \scalef $ and is given by
\begin{equation}
t(\ln \scalef) = \int_{\ln \scalef_0}^{\ln \scalef} \dd \ln \scalef' \, \frac{1}{H(\ln \scalef')}\,.
\end{equation}
Once $ H $ and $ t $ have been self-consistently updated, the differential equation can be solved again.

\section{Details of the SN1987A limits from ALP decay}\label{appendix:sn1987a}
Here we give the full details of our SN1987A likelihood. While a description of the procedure and some of the relevant equations was already provided in the literature~\cite{1009.5714,1702.02964,2104.05727}, we extend the discussion by adding more details on the reference frame transformations and Monte Carlo simulations.

\subsection{Reference frame transformations}
In \cref{fig:illustration}, we show the ALP~$\ax$ emerging from \sn and decaying into two photons~$\gamma_1$ and~$\gamma_2$ with angles $\alpha_1$ and~$\alpha_2$. For convenience, let the $x$-axis be parallel to the direction of the emergent ALP. We denote the ALP rest frame by ``0'' and the lab frame, i.e.\ the rest frame of the SMM satellite, by ``L.'' The Lorentz boost into the lab frame, for dimensionless velocity $\beta$ and $\gamma \equiv 1/\sqrt{1-\beta^2}$, is then given by
\begin{equation}
  \Lambda = \begin{pmatrix}
    \gamma & \beta\,\gamma & 0 & 0\\
    \beta\,\gamma & \gamma & 0 & 0\\
    0 & 0 & 1 & 0\\
    0 & 0 & 0 & 1
  \end{pmatrix} \, .
\end{equation}
Boosting the 4-momentum of an ALP from the rest frame to the lab frame gives
\begin{equation}
p^\mu_{a,\text{0}} = \fourvec{\ma}{0}{0}{0}
\quad \mapsto \quad
p^\mu_{a,\text{L}} = \Lambda^{\mu\nu} \, (p_{a,\text{0}})_\nu = \fourvec{\gamma\,\ma}{\beta\gamma\,\ma}{0}{0} \, .
\end{equation}
The two decay photons are emitted back-to-back in the ALP rest frame, which can be used to boost them into the lab frame:
\begin{equation}
p^\mu_{1/2,\text{0}} = \frac{\ma}{2} \, \fourvec{1}{\pm\cp\,\sth}{\pm\sinp\,\sth}{\pm\cth}
\quad \mapsto \quad
p^\mu_{1/2,\text{L}} = \frac{\gamma\,\ma}{2} \, \fourvec{1 \pm \beta\,\cp\,\sth}{\beta \pm \cp\,\sth}{\pm\frac{1}{\gamma}\sinp\,\sth}{\pm\frac{1}{\gamma}\cth} \, .
\end{equation}
Consequently, the two photon energies~$E_1$ and $E_2$ in the lab frame are
\begin{equation}
  E_{1/2,\text{L}} = \frac{\gamma\,\ma}{2} \left(1 \pm \beta\,\cp\,\sth\right) = \frac{E_a}{2} \left(1 \pm \beta\,\cp\,\sth\right) \, ,
\end{equation}
while the photon angles~$\alpha_1$ and $\alpha_2$ in the lab frame are defined geometrically as the angle between the 3-momenta and the $x$-direction~$\vec{e}_x$ such that
\begin{equation}
  \cos \alpha_{1/2,\text{L}} = \frac{\vec{e}_x \cdot \vec{p}_{1/2}}{|\vec{e}_x| \, |\vec{p}_{1/2}|} = \frac{E_a}{2} \, \frac{\beta \pm \cp\,\sth}{ |\vec{p}_{1/2}|} \, . \label{eq:photonanglesdef}
\end{equation}
Calculating the angles in \cref{eq:photonanglesdef} requires some algebra, in particular we need that
\begin{equation}
  \frac{2}{E_a} \, |\vec{p}_{1/2}| = \left| 1 \pm \beta\, \cp\sth\right| = 1 \pm \beta\, \cp\sth \, , \label{eq:photon_energies}
\end{equation}
where the absolute value in the last line can be dropped because the absolute values of $\beta$, $\cp$, and $\sth$ are all smaller than unity. The angles are therefore given by
\begin{equation}
  \alpha_{1/2,\text{L}} = \arccos \left(\frac{\beta \pm \cp\,\sth}{1 \pm \beta\, \cp\sth} \right) \, . \label{eq:photon_angles}
\end{equation}
Equation~\eqref{eq:photon_angles} reveals additional rotational symmetry in the problem because only the product $\cp\sth$ is important; one could thus e.g.\ set $\theta = \pi$ for simplicity.

\subsection{Monte Carlo procedure}
For each ($\ma$,\,$\gagg$) pair we simulate \num{e7} events.
This numbers reasonable since the {na\"ive} photon fluence in the parameter region of interest is of order \SI{e6}{\cm^{-2}} while the $3\sigma$ fluence limit is about \SI{2}{\cm^{-2}}.
For example, ALPs with $\ma = \SI{1}{\MeV}$ and $\gagg = \SI{e-11}{\GeV^{-1}}$ are just outside of the excluded region; the corresponding acceptance fraction of the MC procedure is about \num{2e-5}, or \num{203 \pm 14} events out of \num{e7}.

We first draw ALP energies from the spectrum in \cref{eq:alp_spectrum_sn1987a} by numerically calculating the inverse of the (normalised) spectrum's cumulative distribution. Similarly, we draw the decay lengths~$L_a$ from an exponential distribution, with mean value parameter given by \cref{eq:alp_decay_length}. For ALPs to yield detectable decays, they have to:
\begin{itemize}
\item[a)] decay at least \SI{e10}{\metre} away from \sn~(the orange shell around \sn in \cref{fig:illustration}), as photons from decays closer to the SN will be absorbed or deflected by its turbulent environment, and
\item[b)] decay before reaching the Earth, as almost no photons will reach the detector beyond this point~(cf.\ discussion in Ref.~\cite{1702.02964}).
\end{itemize}

For photons that satisfy these conditions, we draw a random emission angle for one of the two photons in the ALP rest frame from uniform distributions: $\phi \sim \mathcal{U}(0,\,2\pi)$ and $\cth \sim \mathcal{U}(-1,\,1)$. The angles $\phi$ and $\theta$ of the other photon are the negative of the angles of the first photon, as the two photons are emitted back-to-back. Then, for each of the two photons, we calculate the energies $E_{1/2,\text{L}}$ according to \cref{eq:photon_energies} in the lab frame and check that these energies are within the detector energy window of \SIrange{25}{100}{\MeV}.

Finally, for the photons also making this cut, we calculate the photon angles~$\alpha_{1/2}$ in the lab frame according to \cref{eq:photon_angles}. This allows us to compute the total length of their path, $L_{1/2}$, to the detector and hence their arrival times~$t_{1/2}$, which are~\cite[Sec.~II.C]{1702.02964}
\begin{align}
  L_{\gamma,1/2} &\equiv - L_a \, \cos(\alpha_{1/2}) \sqrt{d_\text{SN}^2 - L_a^2 \sin^2(\alpha_{1/2})} \, , \\
  t_{1/2} &= L_{1/2} = L_a/\beta + L_{\gamma,1/2} - d_\text{SN} \, . \label{eq:arrival_times}
\end{align}

Demanding that photons arrive earlier than \SI{223}{\s} after the arrival of the first neutrino is the last check in the Monte Carlo procedure.

\subsection{Results}

We tabulate the calculated fluence on a logarithmic grid of masses from $\SI{1}{\keV}$ to $\SI{1}{\GeV}$ and ALP-photon couplings from $\SI{e-13}{\GeV^{-1}}$ to $\SI{e-3}{\GeV^{-1}}$. We use a grid spacing of \SI{0.05}{\dex} in both dimensions, leading to a grid of 24,000 points. We chose this grid in order to close the gap between \sn and Horizontal Branch star limits~\cite[e.g.][]{1702.02964}. The resulting table is provided in \gambit as \texttt{DarkBit/data/SN1987A\_DecayFluence.dat}.

The results are shown in \cref{fig:illustration}, where we compare the excluded region from our implementation with that of Ref.~\cite{1702.02964}. Our contours are smoother as we perform full Monte Carlo simulations without the high-mass cut used in Ref.~\cite{1702.02964}. Apart from this difference, we get good qualitative agreement with the previous bounds. However, note that we typically find that our calculated flux is a factor $\lesssim 2$ larger than Ref.~\cite{1702.02964} -- in line with what Ref.~\cite{2109.03244} observe using the non-probabilistic integration methods from Ref.~\cite{1993_Oberauer}. After inspecting the code of Ref.~\cite{1702.02964} and its output, we suspect that apart from the wide binning used in Ref.~\cite{1702.02964} there might also be some ``double counting'' in the implementation of the factors considered in their Eq.~(14).

\subsection{Technical details}
Our scripts, written in Python v3.7.5 and using the \textsf{numpy}, \textsf{scipy}, \textsf{time}, and \textsf{mpi4py} packages, are available at \url{https://github.com/marie-lecroq/ALP-fluence-calculation}. The main script \texttt{functions.py} contains the function \texttt{expected\_photon\_fluence}, which returns the photon fluence~(in units of \si{\cm^{-2}}) given an ALP mass~\texttt{m}~(in units of \si{\eV}) and ALP-photon coupling~\texttt{g}~(in units of \si{\GeV^{-1}}). We also included the script \texttt{run\_analysis\_mpi.py} to run our algorithm in parallel using MPI.


\bibliography{R2}

\end{document}